\documentclass[aps,prl,reprint,superscriptaddress]{revtex4-2}
\usepackage{amsmath}
\usepackage{amssymb}
\usepackage{enumitem}
\usepackage{color}
\usepackage{xcolor}
\usepackage{graphicx} 
\usepackage{dcolumn}  
\usepackage{bm} 
\usepackage[colorlinks=true,linkcolor=blue,citecolor=blue,urlcolor=blue, hypertexnames=false]{hyperref}
\usepackage{feynmp-auto} 

\newcommand{\drop}[1]{}

\definecolor{BluBondi}{rgb}{0.00,0.58,0.71}
\definecolor{tangerine}{rgb}{0.944,0.522,0}
\definecolor{brown}{rgb}{0.633,0.156,0.156}
\definecolor{byzantine}{rgb}{0.74, 0.2, 0.64}

\newcommand{\editor}[2]{%
  \expandafter\newcommand\csname #1note\endcsname[1]{%
    \textcolor{#2}{(\textbf{#1:} \textit{##1})}}%
  \expandafter\newcommand\csname #1\endcsname[1]{%
    \textcolor{#2}{##1}}%
  \expandafter\newcommand\csname #1cancel\endcsname[1]{%
    \textcolor{#2}{\sout{##1}}}%
  \expandafter\newcommand\csname #1change\endcsname[2]{%
    \textcolor{#2}{\sout{##1} ##2}}%
  \newenvironment{#1text}{\color{#2}}{\color{black}}
}
\newcommand{\Trw}{\text{Tr}_\omega}
\newcommand{\TrLn}{\text{Tr}_\omega\text{Ln}}
\newcommand{\ket}[1]{|#1\rangle}
\newcommand{\bra}[1]{\langle #1|}
\newcommand{\braket}[2]{\langle #1 | #2 \rangle}

\newcommand{\Lt}{\text{L}}
\newcommand{\Rt}{\text{R}}
\newcommand{\AllE}{\text{AE}}

\makeatletter
\def\maketitle{
\@author@finish
\title@column\titleblock@produce
\suppressfloats[t]}
\makeatother
\begin{document}
\title{Dynamical pseudopotentials}
%
%
\author{Matteo Quinzi}
\email[corresponding author: ]{matteo.quinzi@epfl.ch}
\affiliation{Theory and Simulation of Materials (THEOS), and National Centre for Computational Design and Discovery of Novel Materials (MARVEL), \'Ecole Polytechnique F\'ed\'erale de Lausanne, 1015 Lausanne, Switzerland}
\author{Tommaso Chiarotti}
\affiliation{Theory and Simulation of Materials (THEOS), and National Centre for Computational Design and Discovery of Novel Materials (MARVEL), \'Ecole Polytechnique F\'ed\'erale de Lausanne, 1015 Lausanne, Switzerland}
\affiliation{Department of Applied Physics and Materials Science, California Institute of Technology, Pasadena, California 91125, USA}
\author{Nicola Marzari}
\affiliation{Theory and Simulation of Materials (THEOS), and National Centre for Computational Design and Discovery of Novel Materials (MARVEL), \'Ecole Polytechnique F\'ed\'erale de Lausanne, 1015 Lausanne, Switzerland}
\affiliation{PSI Center for Scientific Computing, Theory and Data, 5232 Villigen PSI, Switzerland}
\affiliation{Theory of Condensed Matter, Cavendish Laboratory, University of Cambridge, Cambridge, CB3 0US, United Kingdom.}
\date{\today}
%
\begin{abstract}
Pseudopotential theory has greatly driven first-principles calculations in materials, replacing the explicit treatment of the chemically inert core electrons with an effective potential acting only on the valence states. This is inherently an embedding problem, where tracing out the core electrons can be formulated in terms of a dynamical embedding potential. Motivated by this perspective, we first introduce a framework for dynamical (i.e., energy-dependent) pseudopotentials, showing how this leads to generalized norm-conservation conditions. Then, adopting a sum-over-poles representation, we disentangle the number of reference energies from the number of projectors; this allows to reproduce all-electron scattering at many reference energies with great accuracy and over very extended energy ranges.
We further show that these pseudopotentials enter naturally into many-body total energy functionals, leading for the first time to a consistent and unified treatment of the all-electron atom, the pseudo-atom, and the solid within the same electronic-structure theory.

\end{abstract}
\maketitle
%
%
Pseudopotentials are a backbone of modern electronic-structure theory~\cite{PhillipsNew1959, HeineThepseudopotential1970, HamannNormconserving1979, BacheletPseudopotentials1982, KleinmanEfficacious1982, Vanderbilt1985Optimally, TroullerEfficient1990, VanderbiltSoft1990, Rappe1990Optimized, HamannOptimized2013}.
By eliminating the tightly bound and chemically inert core electrons from an explicit treatment and replacing them with a (smoother) effective potential acting solely on the valence manifold, they enabled plane-wave density-functional theory (DFT)~\cite{HohenbergInhomogeneous1964, KohnSelfconsistent1965} as an agile and scalable tool for quantitative materials simulations~\cite{MarzariElectronicstructure2021}.
Formally, integrating out the core electrons is equivalent to an embedding problem: the valence electrons experience an effective energy-dependent potential generated by the degrees of freedom that have been traced out.
This dynamical character was already evident in the early Phillips-Kleinman formulation~\cite{PhillipsNew1959}; 
in spite of this, the pseudopotential framework has evolved predominantly around static approximations.

The construction of pseudopotentials has long been informed by the guiding principle of transferability~\cite{HeineThepseudopotential1970}: the pseudo (PS) atom must reproduce the scattering properties of the all-electron (AE) atom to preserve chemical bonding across different environments.
In the framework of Hamann, Schl\"uter and Chiang~\cite{HamannNormconserving1979}, this requirement is encoded into a set of constraints ensuring that the AE and PS valence eigenvalues agree for a chosen atomic configuration, that their wavefunctions match outside a chosen core radius $R_c$, and that the same total charge is enclosed within $R_c$ (norm conservation).
For a semilocal potential these conditions guarantee the correct logarithmic derivative, and hence the correct scattering properties, in the neighborhood of a target reference energy.
Vanderbilt later introduced two major extensions of this paradigm~\cite{VanderbiltSoft1990}.
First, multiprojector constructions made it possible to target additional reference energies and improve transferability over a broader window.
Second, ultrasoft pseudopotentials (USPP) relaxed norm conservation, compensating the missing charge density through augmentation charges and thereby allowing much softer pseudo-orbitals.

The limitation to static pseudopotentials becomes especially acute for excited-state calculations.
While pseudopotentials often -- but not always -- predict ground-state properties with accuracy comparable to all-electron methods~\cite{BosoniDensityfunctional2024}, their use in excited-state theories can be far more problematic~\cite{OnidaElectronic2002, SchilfgaardeAdequacy2006, GomezInfluence2008, KlimesPredictive2014, GovoniGW1002018}.
Many-body perturbation theory approaches require large numbers of high-lying unoccupied states, whereas conventional pseudopotentials are typically optimized only for valence and a few low-lying excitations.
Transferability is therefore still ultimately limited, especially at higher energies: attempts to include more than two projectors usually run into near-linear dependencies, leading to ill-conditioning and numerical instabilities~\cite{GarrityPseudopotentials2014, vanSettenThePseudoDojo2018}.
In practice, one then resorts either to harder norm-conserving pseudopotentials~\cite{AziziPrecision2025}, with the associated increase in plane-wave cutoffs, or to more elaborate reconstruction schemes such as projector-augmented waves (PAW)~\cite{BlochlProjector1994, KressefromUltrasoft1999, GruneisIonization2014}, with additional algorithmic complexity.
Beyond transferability, the static character of conventional pseudopotentials leads to a different treatment of core and valence electrons in theories where the valence electrons are described by frequency-dependent self-energies, as in GW~\cite{OnidaElectronic2002}, dynamical mean-field theory (DMFT)~\cite{Georges1996Dynamical}, or other beyond-DFT approaches such as dynamical Hubbard~\cite{Chiarotti2024Energies, Chiarotti2025selfconsistent}.
Just as static pseudopotentials enabled the practical large-scale deployment of DFT, dynamical pseudopotentials represent the natural framework to employ when using methods built on dynamical self-energies and spectral functionals.

In this Letter, and motivated by the embedding perspective, we formulate pseudopotential theory as a dynamical embedding problem generated by an auxiliary bath coupled to the valence electrons.
In the presence of an energy-dependent potential, the norm-conservation condition is generalized naturally, yielding augmentation charges directly related to the energy derivative of the pseudopotential.
Unlike early energy-dependent formulations, this construction provides a systematic route to constructing dynamical pseudopotentials that preserve all the transferability requirements of the standard framework and remain compatible with the machinery of modern pseudopotential formulations.
Building on this viewpoint, we represent the pseudopotential in a practical sum-over-poles representation~\cite{Chiarotti2022Unified, Chiarotti2024Energies, Chiarotti2025selfconsistent}, which disentangles the number of reference energies from the number of nonlocal projectors and allows to reproduce all-electron scattering at arbitrarily many reference energies without the numerical instabilities that affect multiprojector schemes.
Last, we further derive a stationary total-energy formulation in which the ionic pseudopotential is screened self-consistently by dynamical augmentation charges.
The resulting framework yields accurate transferability over broad energy windows and places pseudopotentials on the same formal footing as the frequency-dependent self-energies of correlated-electron methods.

The generation of pseudopotentials in a DFT framework starts from an AE atomic calculation, with the Schr\"odinger equation given by $[T~+~v_\text{AE}]\ket{\psi^\text{AE}} = \omega \ket{\psi^\text{AE}}$, where $T$ is the one-particle kinetic energy operator and $v_\text{AE}$ is the screened AE local potential obtained within the chosen approximation for the exchange-correlation functional~\footnote{A braket notation is used even for non-normalizable states, in agreement with Refs.~\cite{VanderbiltSoft1990, HamannOptimized2013}. 
All integrals are still meaningful as they involve wavefunctions up to a finite cutoff radius $R_c$.}.
Spherical symmetry is assumed throughout, and $i$ is a composite index employed for the principal quantum numbers of energy and angular momentum $i=\{n,l,m\}$.
A core radius $R_c$ is then defined and  PS orbitals $\ket{\phi_i}$ are generated for a chosen set of reference energies $\varepsilon_i$~\cite{Hamann1989Generalized}, following the HSC~\cite{HamannNormconserving1979} prescriptions:
(i) eigenergies of the PS orbitals should match the AE ones;
(ii) for distances $r\geq R_c$ from the nucleus the PS wavefunctions should match the AE ones, while they are made smoother inside $R_c$;
and (iii) the logarithmic derivatives of the PS orbitals should also match the AE ones at $R_c$.
The pseudopotential is separated into a local part $v_\text{loc}(r)$ obtained from the pseudization of the AE potential inside the core radius and by a nonlocal part $v_\text{NL}(r_1,r_2)$ vanishing outside $R_c$.
We generalize this construction allowing the nonlocal part to be also energy-dependent, while remaining Hermitian. For a generic energy $\omega$ the nonlinear eigenvalue problem reads:
\begin{equation}
\label{eq:nonlin_eig}
[T + v_\text{loc} + v_\text{PS}(\omega)]\ket{\phi} = \omega \ket{\phi};
\end{equation}
This is in general nonlinear and can be seen as the generalization of the Schr\"odinger equation to a Dyson equation~\cite{Martin_Reining_Ceperley_2016, Chiarotti2022Unified, Chiarotti2024Energies}.
Furthermore, the use of a nonlocal and energy-dependent potential mirrors the structure of the original Phillips-Kleinman construction~\cite{PhillipsNew1959}.
The HSC norm-conservation condition~\cite{HamannNormconserving1979}, ensuring optimal transferability, states that the total PS charge within the core radius must equal the AE one. 
We now show that this condition can be relaxed in the presence of the energy-dependent potential of Eq.~\eqref{eq:nonlin_eig} and that the energy derivatives of the potential take the role of the augmentation charges of the USPP formalism~\cite{VanderbiltSoft1990}.
For a spherically symmetric potential, scattering is fully characterized by the energy-dependent phase shifts of the radial solutions.
These, in turn, are uniquely determined by the logarithmic derivatives of the radial wavefunction at the core radius $R_c$ as a function of energy~\cite{HamannNormconserving1979, Martin_2020}.
Reproducing the scattering properties in the neighborhood of an energy \(\omega\) therefore amounts to reproducing the linear energy dependence of this logarithmic derivative.
Following Shirley and Martin, by manipulating the radial Schr\"odinger equation~\cite{Luders1955Zum, Shirley1989Extended, Shirley1993Many-body} in integro-differential form, one obtains
\begin{equation}
    \label{eq:gncc}
    \partial_\omega\frac{d\,\text{ln}\phi}{dr}\Big|_{r=R_c} = 
    -\frac{2}{|\phi(R_c)|^2}\langle \phi| I - \partial_\omega v_\text{PS}(\omega)|\phi\rangle_{R_c},
\end{equation}
where the logarithmic derivative is evaluated at the cutoff radius \(R_c\).
Equation~\eqref{eq:gncc} shows that, for an energy-dependent pseudopotential, the energy variation of the logarithmic derivative is controlled not only by the norm of the pseudo-wavefunction inside the core, but also by the energy derivative of the pseudopotential itself.
From Eq.~\eqref{eq:gncc} it follows that the pseudopotential is able to reproduce the logarithmic derivative of the AE core for any energy $\omega$ if 
\begin{equation}
    \label{eq:gncc_norm}
    \langle\psi^{\mathrm{AE}}|\psi^{\mathrm{AE}}\rangle_{R_c} = \langle\phi|I - \partial_\omega v_\text{PS}(\omega)|\phi\rangle_{R_c}.
\end{equation}
This last equation generalizes the HSC norm-conservation condition, allowing to construct PS orbitals as soft as possible with dynamical augmentation charges $Q(\omega)=-\bra{\phi}\partial_\omega v_\mathrm{PS}(\omega)|\phi\rangle$ accounting for the charge difference.
The USPP framework can be identified as a particular case of Eq.~\eqref{eq:gncc} where $v_\text{PS}(\omega)$ has a linear energy dependence, as derived from the generalized eigenvalue problem $[T + v_\text{loc}+v_\text{NL}-\omega S]\ket{\phi}=0$ of Ref.~\cite{VanderbiltSoft1990}, where the overlap operator $S$ is built with static augmentation charges.
A result similar to Eq.~\eqref{eq:gncc} is also reported by Shirley and Martin~\cite{Shirley1993Many-body} in the case of relativistic self-energies.
Additionally, Eq.~\eqref{eq:gncc_norm} is reminiscent of the Dyson orbitals normalization condition in Green's function theories, especially in the context of quasiparticle equations~\cite{Martin_Reining_Ceperley_2016, Stefanucci_vanLeeuwen_2013}.
In particular, Eq.~\eqref{eq:gncc} offers a clear picture of the augmentation charges in terms of physical quantities, analogously to the loss of quasiparticle spectral weight in interacting theories.
The generalized norm-conservation condition of Eq.~\eqref{eq:gncc_norm} ensures the transferability of the pseudopotential around an energy $\omega$;
however, it gives no practical prescription on how the pseudopotential must be constructed.
Below, we present an embedding scheme leading to the generation of dynamical pseudopotentials written as sums over poles (SOPs)~\cite{Chiarotti2022Unified, Chiarotti2024Energies} and automatically enforcing Eq.~\eqref{eq:gncc_norm}.
Furthermore, the SOP structure can be exploited to solve exactly the nonlinear eigenvalue problem defined by Eq.~\eqref{eq:nonlin_eig}, by means of the algorithmic-inversion method~\cite{Chiarotti2022Unified, Chiarotti2024Energies, Chiarotti2025selfconsistent}.
%
%

%
The key drive behind the embedding approach employed here is to couple the physical valence sector, described by \(h_0=T+v_\text{loc}\), to an auxiliary bath \(\sigma\) representing the missing core degrees of freedom.
The valence pseudo-electrons can then be viewed as quasiparticles renormalized by the presence of the core, so that their norm in the physical sector alone need not coincide with that of their all-electron counterparts.
Once the two subsystems are coupled, however, they define a closed system \(\mathcal{C}\) in which the total charge is conserved and the usual norm-conservation can be enforced.
Following embedding theory~\cite{Martin_Reining_Ceperley_2016,Stefanucci_vanLeeuwen_2013}, the action of the subsystem $\sigma$ on $h_0$ is represented by a dynamical potential, which in our case is identified as the dynamical pseudopotential $v_\text{PS}(\omega)$ in Eq.~\eqref{eq:nonlin_eig}.
For the chosen set of reference energies, the nonlinear eigenvalue problem of Eq.~\eqref{eq:nonlin_eig} is equivalent to a linear problem $\mathcal{H}\ket{\Psi_i}=\varepsilon_i\ket{\Psi_i}$ in the closed system $\mathcal{C}$ sharing the same eigenvalues~\cite{Martin_Reining_Ceperley_2016, Chiarotti2024Energies, Ferretti2024Greens}
\begin{equation}
    \label{eq:embedding_ham}
    \begin{pmatrix}
        T + v_\text{loc}  &   V  \\
        \overline{V}^\dagger & \sigma
    \end{pmatrix}
    \begin{pmatrix}
        \ket{\phi_i} \\
        \ket{q_i}
    \end{pmatrix}
    = \varepsilon_i
    \begin{pmatrix}
        \ket{\phi_i} \\
        \ket{q_i}
    \end{pmatrix},
\end{equation}
where we have introduced couplings $V$ and $\overline{V}^\dagger$ between the two subsystems.
While the pseudopotential has to be a Hermitian operator in the valence sector, for a generic nonlinear problem the Hamiltonian of Eq.~\eqref{eq:embedding_ham} can be non-Hermitian, with $V\neq\overline{V}$, resulting in different right and left eigenvectors $\{\ket{\Psi^\text{R/L}_i}\}$.
The conditions for the pseudopotential to be Hermitian in presence of a non-Hermitian bath are discussed in the End Matter.
We partition the eigenvectors into a PS orbital component $\ket{\phi_i}$ and into an augmentation component $\ket{q_i}$.
Since the radial PS orbitals are always generated as real valued, the non-Hermiticity of the problem may result in having different right $\ket{q_i^\text{R}}$ and left $\ket{q_i^\text{L}}$ eigenvector components.
We enforce the HSC norm-conservation condition~\cite{HamannNormconserving1979} in the closed system $\mathcal{C}$:
\begin{equation}
    \label{eq:ncc_emb}
    \braket{\Psi_i^\Lt}{\Psi_j^\Rt} = \braket{\phi_i}{\phi_j} + \braket{q_i^\Lt}{q_j^\Rt} = \braket{\psi^{\AllE}_i}{\psi^{\AllE}_j}
\end{equation}
with the dynamical pseudopotential obtained as the embedding potential acting on the $h_0$ subsystem due to the presence of the bath $\sigma$:
\begin{equation}
    \label{eq:dyn_pseudo_sop}
    v_\text{PS}(\omega) = V \,(\omega I - \sigma )^{-1}\,\overline{V}^\dagger.
\end{equation}
If Eq.~\eqref{eq:ncc_emb} holds, then the generalized norm-conservation condition of Eq.~\eqref{eq:gncc_norm} is enforced on the soft PS orbitals in the physical system.
To show this, we first define the augmentation charges of the system as $Q_{ij}~:=~\braket{\psi^{\AllE}_i}{\psi^{\AllE}_j}_{R_c} - \braket{\phi_i}{\phi_j}_{R_c}=\braket{q_i^\Lt}{q_j^\Rt}$.
It is then straightforward to show that Eq.~\eqref{eq:embedding_ham} ensures that the augmentation charges are directly related to the energy derivative of the pseudopotential of Eq.~\eqref{eq:dyn_pseudo_sop}, since $\ket{q_i^\Rt}=(\varepsilon_i - \sigma)^{-1}\overline{V}^\dagger\ket{\phi_i}$, with an analogous relation for $\bra{q_i^L}$, giving
\begin{equation}
    \label{eq:dyn_aug_charges}
    Q_{ij} = \braket{q_i^\Lt}{q_j^\Rt} = \bra{\phi_j} V (\varepsilon_j - \sigma)^{-1}(\varepsilon_i - \sigma)^{-1}\overline{V}^\dagger\ket{\phi_i},
\end{equation}
which is exactly $Q(\varepsilon_i)=-\bra{\phi_i}\partial_\omega v_\text{PS}(\varepsilon_i)\ket{\phi_i}$ for the diagonal matrix elements.
Furthermore, Eq.~\eqref{eq:dyn_aug_charges} clearly characterizes the degrees of freedom of the bath $\sigma$ as augmentation charges.
Although the present pseudopotential is genuinely energy-dependent, the fundamental objects entering its construction arise in direct analogy with the USPP formalism from the linear embedding problem of Eq.~\eqref{eq:embedding_ham}.
Adopting the notation of Ref.~\cite{VanderbiltSoft1990}, we introduce the projectors
\begin{align}
    \label{eq:chi}
    \ket{\chi_i} &= (\varepsilon_i - h_0)\ket{\phi_i} = V\ket{q_i^\mathrm{R}}
\end{align}
which are localized within the cutoff radius~\cite{VanderbiltSoft1990}, since the PS orbitals and pseudopotential coincide with the AE ones outside \(R_c\).

\begin{figure*}[t]
    \centering
    \includegraphics[width=0.49\textwidth]{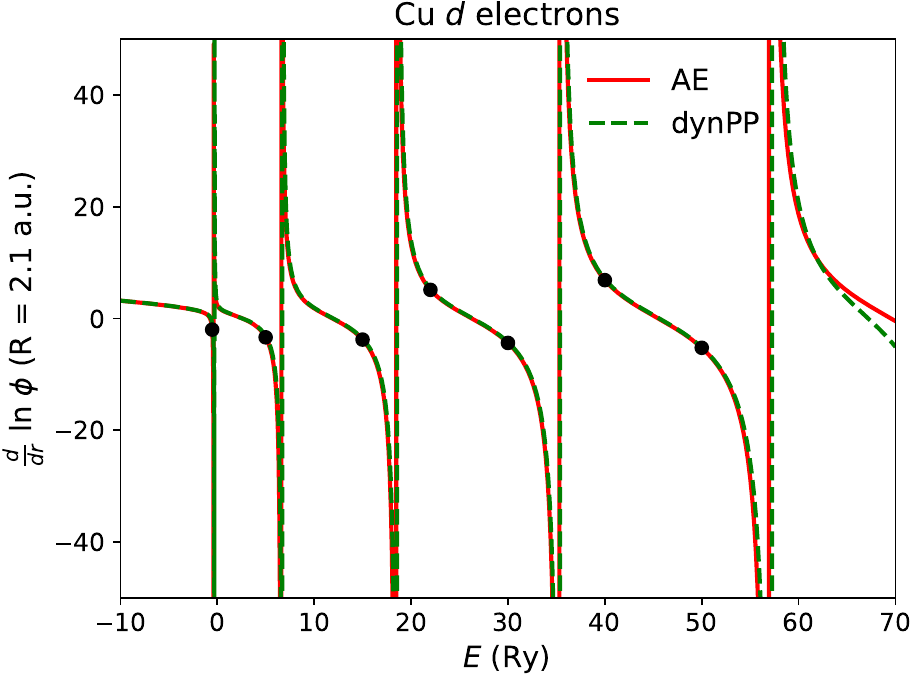}
    \hfill
    \includegraphics[width=0.49\textwidth]{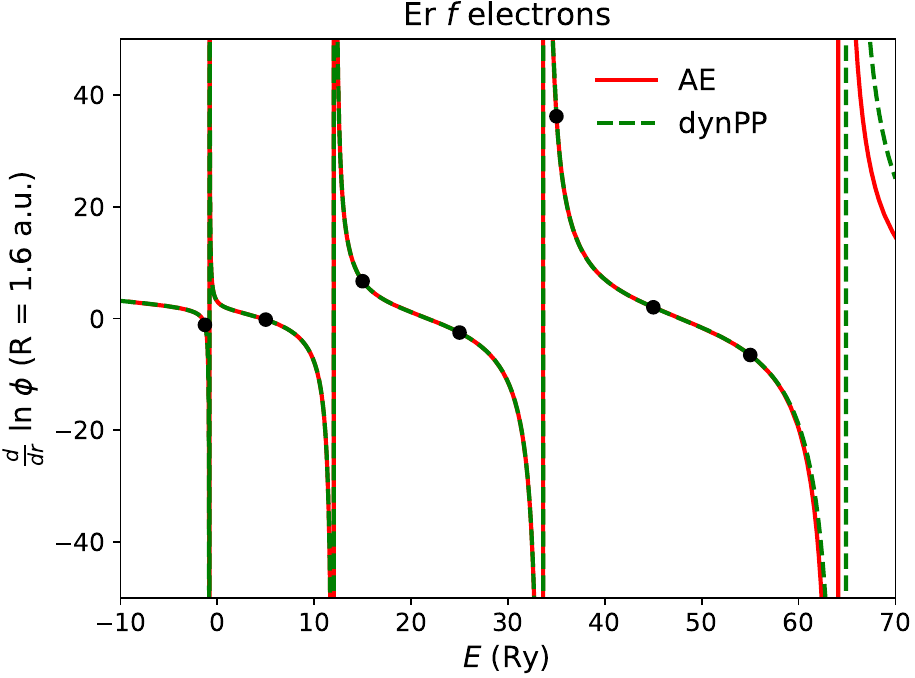}
    \caption{
    Logarithmic derivative as a function of energy for the all-electron (AE) potential, and dynamical pseudopotential (dynPP) for the (left) copper $d$ electrons at distance $R=2.1$ (a.u.) and (right) erbium $f$ electrons at distance $R=1.6$ (a.u.).
    The reference energies are marked with black dots. In both cases the pseudopotential is built including up to seven reference energies, while only three projectors are used following the truncation strategy.
    }
    \label{fig:lders}
\end{figure*}

The dynamical pseudopotential is built with a sum-over-poles representation, in terms of scalar poles $\Omega_s$ and matrix valued residues $\Gamma^s_{\alpha\beta}$$=V_{\alpha s}\overline{V}^{\dagger}_{s\beta}$, where the spatial dependence is given by a set of ortho-normal basis states \(\{\ket{b_\alpha}\}\) localized within the core radius: 
\begin{equation}
    \label{eq:dyn_pseudo_generated}
    v_\text{PS}(\omega) = \sum_{\alpha\beta}\sum_s \ket{b_\alpha} \frac{V_{\alpha s}\overline{V}^\dagger_{s\beta}}{\omega - \Omega_s}\bra{b_\beta},
\end{equation}
with the only requirement for the localized basis to be complete in the space spanned by the (normalized) \(\{\ket{\chi_i}\}\) projectors, since Eq.~\eqref{eq:nonlin_eig} is satisfied whenever \(v_{\mathrm{PS}}(\varepsilon_i)\ket{\phi_i}=\ket{\chi_i}\).
A crucial advantage of the present SOP construction is that the number of poles and the number of basis functions are not tied to each other.
While the number of poles is fixed by the number of reference energies chosen, the nonlocal part of the pseudopotential can be represented with a smaller basis, provided that this basis accurately spans the space of the (normalized) projectors \(\{\ket{\chi_i}\}\).
To quantify the completeness of a localized orthonormal basis \(\{\ket{b_\alpha}\}\) in the \(\{\ket{\chi_i}\}\) space, we introduce the functional
\begin{align}
\label{eq:chi_spread_functional}
\Omega_\chi[\{b_k\}]&=\frac{1}{N}\sum_{i=1}^{N}\Bigg[\braket{\chi_i}{\chi_i}-\sum_{\alpha=1}^{N_b}|\braket{\chi_i}{b_\alpha}|^2\Bigg]\nonumber\\
&\quad +\sum_{\alpha=1}^{N_b}\lambda_\alpha(\braket{b_\alpha}{b_\alpha}-1),
\end{align}
where \(N\) is the number of \(\{\ket{\chi_i}\}\) projectors, \(N_b\) is the number of basis states, and \(\lambda_\alpha\) are Lagrange multipliers enforcing orthonormality.
For an orthonormal basis, \(\Omega_\chi=0\) for a complete representation of the \(\{\ket{\chi_i}\}\) space, while \(\Omega_\chi=1\) for a basis that does not project onto it at all.
Assuming that the basis states are a linear combination of the $\{\ket{\chi_i}\}$, the minimization of Eq.~\eqref{eq:chi_spread_functional} yields
\begin{equation}
\label{eq:variational_basis}
\ket{b_\alpha}=\lambda_\alpha^{-1/2}\sum_{i=1}^N \ket{\chi_i}\Lambda_{i\alpha},
\end{equation}
where \(\lambda_\alpha\) and \(\Lambda_{i\alpha}\) are, respectively, the eigenvalues and eigenvector components of the overlap matrix \(S_{ij}=\braket{\chi_i}{\chi_j}\).
The structure of Eq.~\eqref{eq:chi_spread_functional} is reminiscent of the spread functional introduced for maximally localized Wannier functions~\cite{Marzari1997Maximally}.
This variational approach immediately provides a controlled truncation strategy: one retains only the eigenvectors of \(S_{ij}\) with eigenvalues above a chosen threshold. In this way, the basis is compressed while maintaining completeness in the relevant projector space. 
The residual incompleteness is measured directly by Eq.~\eqref{eq:chi_spread_functional}, which for this optimal truncated basis is proportional to the sum of the discarded eigenvalues of \(S_{ij}\).
In contrast, conventional multiprojector approaches~\cite{VanderbiltSoft1990, HamannOptimized2013} impose the orthonormality constraint \(\braket{b_\alpha}{\phi_i}=\delta_{\alpha i}\), which forces the number of basis states to equal the number of reference energies and can therefore lead to numerical instabilities when the \(\{\ket{\chi_i}\}\) become nearly linearly dependent, even for distinct \(\varepsilon_i\).
At variance with static pseudopotentials, though, the solutions of the nonlinear problem in Eq.~\eqref{eq:nonlin_eig} will be in general nonorthogonal between each others, while bi-orthogonality can always be guaranteed in the closed system $\mathcal{C}$ in Eq.~\eqref{eq:embedding_ham}.
As a representative case, we assess the transferability of the present SOP dynamical pseudopotential by showing the logarithmic derivative for the copper $d$ electrons and for the erbium $f$ electrons over a wide energy range as reported in Fig.~\ref{fig:lders} for the PBE exchange-correlation functional~\cite{PBE_PhysRevLett.77.3865_1996}.
The valence configurations used in the generation are $3s^2\,3p^{6}\,3d^{9.5}\,4s^{1.5}$, with a core radius of $R_c=2.0$ (a.u.) in the $d$-channel for copper and $5s^2\,5p^6\,4f^{12}$, with a core radius of $R_c=1.5$ (a.u.) in the $f$-channel for erbium.
In both cases the local potential is obtained from a polynomial pseudization of the AE one.
The logarithmic derivative is evaluated numerically by outward integration of the radial Schr\"odinger equation.
We consider up to seven reference energies in the construction of the pseudopotentials, starting from the $3d$ bound state for copper and from the $4f$ bound state for erbium and including additional references every time the agreement between PS and AE logarithmic derivatives deteriorates, up to $50$ Ry for copper and $55$ Ry for erbium.
We construct the localized basis set with the normalized eigenvectors of the overlap matrix $S_{ij}$, truncating all states with an eigenvalue lower than $10^{-5}$.
In both cases, this procedure results in having only three basis states, out of seven reference energies, highlighting the nearly linear dependence of the $\ket{\chi_i}$ projectors.
These dynamical pseudopotentials reproduce the scattering properties of the AE atom up to energies of $60$ Ry, well above the requirements for excited-state properties, typically around $30$ Ry~\cite{KlimesPredictive2014}.
The present results establish a clear advantage compared to other multiprojector pseudopotentials~\cite{VanderbiltSoft1990, HamannOptimized2013}, where the inclusion of more reference states would result in an equal amount of projectors -- computationally more expensive and usually numerically ill-conditioned.
In passing, we emphasize that the energy dependence of the potential does not solve the problem of introducing ghost states, which is ultimately related to the separable structure of the potential~\cite{Gonze1991Analysis} and is typically dealt with by adjusting the pseudization of the local potential $v_\text{loc}$ or varying the number of reference energies, and also tested a posteriori~\cite{Prandini2018Precision}.
%
%
%

%
Finally, the pseudopotential needs to be unscreened in order to remove the Hartree and exchange-correlation contribution coming from the valence electrons of the isolated atom, delivering a ionic pseudopotential ready to be used in the target solid state calculation.
This unscreening procedure is typically dealt with by introducing a total energy functional of the ionic pseudopotential~\cite{VanderbiltSoft1990,Louie1982Nonlinear} and comparing its functional derivative with Eq.~\eqref{eq:nonlin_eig}.
We outline a consistent variational framework for total energies and self-consistency in presence of a dynamical pseudopotential.
Since energy-dependent potentials are most naturally described at the level of the one-particle Green's function \(G(\omega)\), this framework is provided by Green's function functionals such as those of Luttinger-Ward and Klein~\cite{Luttinger1960Ground-state, Klein1961Perturbation, Baym1961Conservation, Almbladh1999Variational}.
We therefore embed the present dynamical pseudopotentials within Kohn-Sham DFT~\cite{HohenbergInhomogeneous1964, KohnSelfconsistent1965} by deriving a total-energy expression in the form of a Klein functional in the presence of an embedding potential, following Refs.~\cite{Ferretti2024Greens, Ferretti2025Functional}.
This essential step provides the stationary formulation needed to treat the pseudopotential and its associated augmentation sector self-consistently, and places the ionic pseudopotential on the same formal footing as the dynamical self-energies appearing in correlated-electron methods.
The resulting total-energy functional reads
\begin{align}
    \nonumber
    \label{eq:dynamical_total_energy_functional}
    E_K[G_0,G] =& \Trw[v^\text{ion}_\text{PS}(\omega) G(\omega)] + E_\text{H}[n] + E_\text{xc}[n+n_c] \\
    \nonumber 
    & + \TrLn[G_0^{-1}G(\omega)] \\
    \nonumber
    & + \Trw[I-(G_0^{-1}+v^\text{ion}_\text{loc})G(\omega)]\\
    & + \Trw[h_0 G_0(\omega)] + \int \, dr\, v^\text{ion}_\text{loc}\,(r)n(r),
\end{align}
where \(E_\text{H}\) is the Hartree energy, \(E_\text{xc}\) is an approximation to the exchange-correlation functional, and \(v^\text{ion}_\text{loc}\) and \(v^\text{ion}_\text{PS}(\omega)\) denote the local and dynamical ionic pseudopotentials, respectively.
A nonlinear frozen core correction $n_c$ can be included in $E_\text{xc}$ to improve the transferability~\cite{Louie1982Nonlinear} when the electronic configuration differs from the reference one used in the generation of the pseudopotential. 
The noninteracting Green's function is $G_0^{-1}(\omega)=\omega I - T - v_\text{loc}^\text{ion}$, while the interacting Green's function is defined according to Eq.~\eqref{eq:nonlin_eig} as $G^{-1}(\omega)=\omega I - T-v_\text{loc}-v_\text{PS}(\omega)$.
The charge density $n(r)$ to be used within the functional of Eq.~\eqref{eq:dynamical_total_energy_functional} is the physical charge density reconstructed within the embedding construction in Eq.~\eqref{eq:embedding_ham}.
It is then convenient to introduce a kernel operator $K(r,\omega)$ that reconstructs the charge density within the cutoff radius, ensuring charge-conservation:
\begin{equation}
    \label{eq:kernel}
    K(r,\omega) = \frac{\delta n(r)}{\delta G(\omega)}=\ket{r}\bra{r} - f(r,\omega) \partial_\omega v_\text{PS}(\omega),
\end{equation}
where $f(r,\omega)$ is an energy-dependent function that modulates the shape of the charge density, and can be used either to further pseudize the augmentation charges, or to reconstruct the AE charge density of the isolated atom at any given energy and normalized as $\int_0^{R_c}f(r,\omega)=1$ for each value of $\omega$.
The definition of the kernel in Eq.~\eqref{eq:kernel} is consistent with the one given in Ref.~\cite{Laasonen1991Implementation}: in the USPP case $K(r)$ is energy-independent and the product $-f(r)\partial_\omega v_\text{PS}$ returns the usual augmentation charges, while in the case of norm-conserving pseudopotentials it is simply given by $\ket{r}\bra{r}$.
We then introduce the spectral density as 
\begin{equation}
    \label{eq:spectral_rho}
    \rho(r,\omega) = \sum_s \delta(\omega - \varepsilon_s) \bra{\phi_s}K(r,\varepsilon_s)\ket{\phi_s},
\end{equation}
which generalizes the reconstruction of the charge density of the USPP formalism~\cite{VanderbiltSoft1990, Laasonen1991Implementation, Laasonen1993Car-parrinello} to the case of dynamical augmentation charges.
The charge density is readily obtained as $n(r)=\int_{-\infty}^\mu d\omega\, \rho(r,\omega)$, with $\mu$ the chemical potential of the system.
The Dyson equation is derived from the stationarization of the functional in Eq.~\eqref{eq:dynamical_total_energy_functional} with respect to $G(\omega)$:
\begin{align}
    \label{eq:dyson_from_functional}
    \nonumber 
    G_0^{-1}(\omega) - G^{-1}(\omega) =& v_\text{PS}^\text{ion}(\omega)
    +\int dr\, v_\text{Hxc}(r)K(r,\omega)\\ 
    &+\int dr\, v_\text{loc}^\text{ion}(r)\,\big[K(r,\omega)-\ket{r}\bra{r}\big],
\end{align}
with $v_\text{Hxc}(r)=\frac{\delta (E_\text{H}+E_\text{xc})}{\delta n(r)}$.
The ionic potentials are obtained by unscreening the potentials of Eq.~\eqref{eq:nonlin_eig} according to
\begin{align}
    \label{eq:descreening_loc}
    v_\text{loc} &= v_\text{loc}^\text{ion} + v_\text{Hxc}\\
    \label{eq:descreening_ps}
    v_\text{PS}(\omega) &= v_\text{PS}^\text{ion}(\omega) +\int dr\, v_\text{loc}(r)\, \big[K(r,\omega)-\ket{r}\bra{r}\big].
\end{align}
The (un)screening of the potentials in the USPP case~\cite{VanderbiltSoft1990, Laasonen1991Implementation, Laasonen1993Car-parrinello} is recovered if a linear dependence on the energy is considered.
The kernel of Eq.~\eqref{eq:descreening_ps} can be either kept fixed in the atomic configuration or updated self-consistently with $v_\mathrm{PS}(\omega)$ according to Eq.~\eqref{eq:kernel}, achieving additional flexibility.
Within the SOP formalism, the (un)screening can be carried out by updating the residues of the potential, while keeping the poles fixed.
The potential in Eq.~\eqref{eq:descreening_ps} is now self-consistently screened with the augmentation charges which are updated according to the change in the eigenvalues and eigenvectors of Eq.~\eqref{eq:nonlin_eig}, at variance with the USPP framework~\cite{VanderbiltSoft1990}, where augmentation charges are frozen in the atomic configuration.
This additional flexibility can improve transferability when the electronic valence configuration of the target system is sensibly different from the one used in the generation of the pseudopotential.
Beyond their natural role in Green's function functionals, the present dynamical pseudopotentials can already be employed at the DFT level.
Using the algorithmic-inversion method, the nonlinear eigenvalue problem of Eq.~(\ref{eq:nonlin_eig}), representing a  full dynamical treatment, can be mapped onto a static Hamiltonian carrying additional auxiliary degrees of freedom; this can then be solved with standard diagonalization techniques.
The associated computational overhead is modest: if \(h_0\) is represented in a given basis, for example plane waves, by an \(N\times N\) matrix, the linearized problem has dimension only \(N+N_p\), where \(N_p\) is the number of poles in the pseudopotential. Thus, each pole introduces only one auxiliary degree of freedom, making the dynamical extension a relatively lightweight generalization of standard DFT implementations.
%
%
%

%
In conclusion, we have reformulated the problem of valence electrons in the field of core electrons as an embedding problem, giving rise to dynamical (i.e., energy dependent) pseudopotentials. This leads to generalized norm-conservation conditions and dynamical augmentation charges linked to the energy derivative of the pseudopotential, in analogy to the loss of quasiparticle spectral weight in interacting theory~\cite{Martin_Reining_Ceperley_2016}; the well established USPP framework~\cite{VanderbiltSoft1990} corresponds to the particular case where the pseudopotential has a linear dependence on the energy. Second, once a sum-over-poles~\cite{Chiarotti2022Unified, Chiarotti2024Energies} formulation for the pseudopotential is adopted, one can show that logarithmic derivatives and scattering properties of the all-electron atom and the pseudo-atom can be matched across extremely wide energy ranges --- thanks to the disentanglement of the number of reference energies from the number of projectors --- leading to systematic transferability well beyond the reach of current schemes. Third, dynamical pseudopotentials mark a shift in perspective from static theories to frequency-dependent ones, that allows for a unified treatment --- within a given many-body functional --- of the all-electron atom, the pseudo-atom, and the molecule or the solid, and providing a consistent correlated-electron treatment at the level of many-body perturbation theory/GW~\cite{OnidaElectronic2002}, dynamical mean-field theory~\cite{Georges1996Dynamical}, and dynamical Hubbard functionals~\cite{Chiarotti2024Energies, Chiarotti2025selfconsistent}. 

We acknowledge fruitful and stimulating discussions with A.~Ferretti.
The uspp-736 code from D.~Vanderbilt has been used to perform the all-electron calculations and to generate the atomic orbitals.
This work was supported by the Swiss National Science Foundation (SNSF) through Grant No.213082 (M.Q., T.C.) and NCCR MARVEL (N.M.), a National Centre of Competence in Research through Grant No.205602.
\appendix
\section{END MATTER}
\label{sec:end-matter}
\subsection{Practical construction of sum-over poles pseudopotentials.}
We describe here the practical procedure used to construct ab initio dynamical SOP pseudopotentials.
The required inputs are the radial AE and pseudo PS orbitals, together with the local potential $v_\text{loc}$ obtained by pseudizing the AE potential inside the core radius $R_c$.
From these quantities, the embedding problem of Eq.~\eqref{eq:embedding_ham} is solved by determining the bath $\sigma$ and the couplings $V$ and $\overline V^\dagger$.
The embedding problem of Eq.~\eqref{eq:embedding_ham} defines the following linear system of equations:
\begin{align}
   h_0 \ket{\phi_i} + V\ket{q_i^\mathrm{R}} =& \varepsilon_i \ket{\phi_i}\\
   \overline{V}^\dagger\ket{\phi_i} +\sigma\ket{q_i^\mathrm{R}} =&\varepsilon_i \ket{q_i^\mathrm{R}}\\
   \bra{\phi_i}h_0 + \bra{q_i^\mathrm{L}}\overline{V}^\dagger =&\varepsilon_i\bra{\phi_i}\\
   \bra{\phi_i}V + \bra{q_i^\mathrm{L}}\sigma =& \varepsilon_i\bra{q_i^\mathrm{L}},
\end{align}
where, in the case of a non-Hermitian bath, we consider both left and right eigenvectors, while imposing that the PS orbital components are identical in the two cases.
With the PS orbitals and the local potential $v_\text{loc}$, one first constructs the projectors $\ket{\chi_i}$ of Eq.~\eqref{eq:chi} and then evaluates the matrices $B_{ij}$ and $B^\dagger_{ij}$,
\begin{align}
    \label{eq:B_mat}
    B_{ij} &= \braket{\phi_i}{\chi_j} = \bra{q_i^\mathrm{L}}(\varepsilon_i-\sigma)\ket{q_j^\mathrm{R}},\\
    \label{eq:Bdag_mat}
    B^\dagger_{ij} &= \braket{\chi_i}{\phi_j} = \bra{q_i^\mathrm{L}}(\varepsilon_j-\sigma)\ket{q_j^\mathrm{R}}.
\end{align}
The augmentation charges are computed from the difference between the AE and PS overlaps inside the core region,
\begin{equation}
    Q_{ij}=\braket{\psi^{\mathrm{AE}}_i}{\psi^{\mathrm{AE}}_j}_{R_c}-\braket{\phi_i}{\phi_j}_{R_c}.
\end{equation}
In passing, we stress that the familiar USPP relation linking the Hermiticity of the matrix $B_{ij}$ to the augmentation charges $Q_{ij}$ follows directly from the embedding scheme:
\begin{align}
\label{eq:hermiticity_BQ}
    B_{ij} - B^\dagger_{ij} = (\varepsilon_i - \varepsilon_j)Q_{ij}.
\end{align}
In scalar-relativistic calculations, Eq.~\eqref{eq:hermiticity_BQ} is generally violated because the AE Hamiltonian contains relativistic corrections that are absent from the operator $T+v_{\mathrm{loc}}$~\cite{HamannOptimized2013}.
In these cases, we enforce Eq.~\eqref{eq:hermiticity_BQ} for the off-diagonal elements of $Q_{ij}$.
Once these intermediate quantities are available, the construction reduces to determining the augmentation-sector components $\ket{q_i^{\mathrm{R/L}}}$, the bath Hamiltonian $\sigma$, and the couplings $V$ and $\overline{V}^{\dagger}$.

First, the augmentation-sector components $\ket{q_i^{\mathrm{R/L}}}$ are introduced by requiring that their overlap reproduces the augmentation matrix $Q_{ij}=\braket{q_i^\mathrm{L}}{q_j^\mathrm{R}}$.
A convenient explicit realization is obtained through a square-root factorization of $Q_{ij}$:
\begin{align}
    \label{eq:aug_charge_decomposition}
    \braket{p}{q_i^\Rt} &= \Tilde{Q}^{1/2}_{pp} U_{pi}\\
    \braket{q_j^\Lt}{p} &= U^\dagger_{jp} \Tilde{Q}^{1/2}_{pp},
\end{align}
where the index $p$ labels a basis for the bath $\sigma$, $U_{pi}$ is the matrix containing the components of the eigenvectors of $Q_{ij}$, and $\Tilde{Q}_{pp}$ is the diagonal matrix of its eigenvalues.
This factorization of $Q_{ij}$ is always possible since the matrix is Hermitian.
As a shorthand notation, we introduce the matrices $q^{\mathrm{R}}_{pi}=\braket{p}{q_i^{\mathrm{R}}}$ and $q^{\mathrm{L}\dagger}_{ip}=\braket{q_i^{\mathrm{L}}}{p}$.
Additionally, if the matrix of the augmentation charges $Q_{ij}$ is positive semi-definite, then its eigenvalues are all real-valued and non-negative and the matrices $q^{\mathrm{R}}_{pi}$ and $q^{\mathrm{L}\dagger}_{ip}$ are adjoint to one another.

Second, the bath Hamiltonian $\sigma$ is reconstructed in the basis $\ket{p}$ as 
\begin{equation}
    \label{eq:bath_from_charges}
    \sigma_{p_1 p_2} = \sum_{ji}\left(q^{L\dagger}\right)^{-1}_{p_1 j}\left( Q_{ji}\varepsilon_i -  B^\dagger_{ji}\right) \left(q^R\right)^{-1}_{i p_2},
\end{equation}
where the matrix $Q_{ji}\varepsilon_i - B^\dagger_{ji}$ is Hermitian, following Eq.~\eqref{eq:hermiticity_BQ}.
The eigenvalues of the bath $\sigma$ are the poles of the dynamical pseudopotential.
If the matrices $q^\mathrm{R}_{pi}$ and $q^{\mathrm{L}\dagger}_{ip}$ are adjoint to one another, then $\sigma$ is Hermitian, following Eq.~\eqref{eq:bath_from_charges}, and its eigenvalues are real valued.
Physically, this corresponds to the case where the norm of the PS orbitals is always smaller than that of the AE ones.
In the general case, the left and right  eigenvectors $\ket{s^{\Rt/\Lt}}$ of $\sigma$ form a bi-orthogonal basis in the augmentation sector.

Finally, the couplings $V$ and $\overline V^\dagger$ are expanded in the localized basis $\ket{b_\alpha}$ of Eq.~\eqref{eq:variational_basis} as $V = \sum_{\alpha s} \ket{b_\alpha} V_{\alpha s} \bra{s^\mathrm{L}}$ and $\overline{V}^\dagger = \sum_{s\alpha} \ket{s^\mathrm{R}} \overline{V}_{s\alpha}^\dagger \bra{b_\alpha}$.
The matrix elements of the couplings are computed as:
\begin{align}
    \label{eq:coupling}
    V_{\alpha s} &= \sum_j \braket{b_\alpha}{\chi_j} \left(q^\Rt\right)^{-1}_{js}\\
    \label{eq:coupling_bar}
    \overline{V}^\dagger_{s\alpha} &= \sum_j (q^{\Lt\dagger})^{-1}_{sj}\braket{\chi_j}{b_\alpha},
\end{align}
where the components of $\ket{q_i^\text{R/L}}$ have been transformed to the bath eigenbasis $\ket{s^\text{R/L}}$ as $q^{\mathrm{R}}_{si}=\braket{s^{\mathrm{L}}}{q^{\mathrm{R}}_i}$ and $q^{\mathrm{L}\dagger}_{is}=\braket{q_i^{\mathrm{L}}}{s^{\mathrm{R}}}$.
If the matrices $q^{\mathrm{R}}_{si}$ and $q^{\mathrm{L}\dagger}_{is}$ are adjoint to one another, then the coupling matrices $V_{\alpha s}$ and $\overline{V}^{\dagger}_{s\alpha}$ are also adjoint to one another, following Eq.~\eqref{eq:coupling} and Eq.~\eqref{eq:coupling_bar}.
The requirement that the pseudopotential be Hermitian for any real-valued $\omega$ is compatible either with a SOP structure having real poles and Hermitian residues, which follows when the augmentation matrix $Q_{ij}$ is positive semidefinite, or with pairs of complex-conjugate poles accompanied by complex-conjugate residues.
Both cases can occur in practice.
In particular, in the cases studied here, the use of real-valued pseudo-orbitals, together with the fact that the eigenvalues of $Q_{ij}$ are always real, has been a sufficient condition to generate a dynamical pseudopotential that is Hermitian when evaluated for real energies.
%

%
%
%
\renewcommand\thefigure{S\arabic{figure}}
\renewcommand\thesection{S\arabic{section}}
\renewcommand\theequation{S\arabic{equation}}
\renewcommand\thetable{S\arabic{table}}
\setcounter{section}{0}
\setcounter{figure}{0}
\setcounter{equation}{0}
\setcounter{table}{0}
\def\appendixname{}
%
%
\clearpage
\title{Supplemental material for \\
``Dynamical pseudopotentials''}
\maketitle
\onecolumngrid
%
%
%
%
\section{Derivation of the generalized norm-conservation condition}
We give here the detailed derivation of the generalization of the HSC~\cite{Hamann1989GeneralizedSupp} norm-conservation condition to the case of a dynamical pseudopotential.
Our derivation follows the one given by Shirley~\cite{Shirley1989ExtendedSupp} and Martin~\cite{Martin_2020Supp}, based on Ref.~\cite{Luders1955ZumSupp}.
The most general case is obtained for a nonlocal and energy-dependent pseudopotential $v_{PS}(r_1,r_2;\varepsilon)$.
In the following we will also assume that the pseudopotential is a Hermitian operator for any real-valued energy $\varepsilon$, so that the solution of the radial equation can be taken as real-valued for any given energy.
Assuming spherical symmetry, the Schr\"odinger equation for the pseudo (PS) orbital $\phi(r)$ reads (in atomic units)
\begin{equation}
    \label{eq:ks_eq_dyn_pseudo}
         -\frac{1}{2} \frac{d^2}{dr_1^2}\phi(r_1,\varepsilon) + \frac{l(l+1)}{2r_1^2}\phi(r_1,\varepsilon) + \int\,dr_2 v_\text{PS}(r_1,r_2,\varepsilon)\phi(r_2,\varepsilon) = \varepsilon \phi(r_1,\varepsilon)
\end{equation}
We start introducing the following change of variable
\begin{equation}
    \label{eq:var_x}
    x(r,\varepsilon) := \frac{d}{dr}\text{ln}\phi(r,\varepsilon) = \frac{\phi'(r,\varepsilon)}{\phi(r,\varepsilon)},
\end{equation}
where $\phi'$ indicates that we have taken the derivative with respect to $r$.
Then 
\begin{equation}
    \label{eq:var_x'}
    x' = \frac{\phi''}{\phi} - \left(\frac{\phi'}{\phi^2}\right)^2 = \frac{\phi''}{\phi} - x^2.
\end{equation}
We can then substitute $\phi''$ appearing in Eq.~\eqref{eq:ks_eq_dyn_pseudo} with Eq.~\eqref{eq:var_x'}
\begin{align}
\label{eq:change_phi_with_x}
    &-\frac{1}{2}\left[x'(r_1,\varepsilon) + x^2(r_1,\varepsilon)\right]\phi(r_1,\varepsilon) + \frac{l(l+1)}{2r_1^2}\phi(r_1,\varepsilon) + \int_0^\infty \,dr_2 \, v_\text{PS}(r_1,r_2;\varepsilon)\phi(r_2,\varepsilon) = \varepsilon \phi(r_1,\varepsilon)
\end{align}
and we divide by $\phi(r_1,\varepsilon)$, assuming it is different from zero
\begin{align}
\label{eq:divide_by_phi}
    &x'(r_1,\varepsilon) + x^2(r_1,\varepsilon) - \frac{l(l+1)}{r_1^2} = -2\left[\varepsilon - \int_0^\infty \, dr_2 \, v_\text{PS}(r_1,r_2;\varepsilon) \frac{\phi(r_2,\varepsilon)}{\phi(r_1,\varepsilon)}\right].
\end{align}
We now take the derivative with respect to energy of Eq.~\eqref{eq:divide_by_phi}
\begin{align}
    &\frac{d}{dr}\frac{\partial x(r_1,\varepsilon)}{\partial \varepsilon} + 2 x(r_1,\varepsilon)\frac{\partial x(r_1,\varepsilon)}{\partial \varepsilon} =\\
\nonumber
    & = -2 \left\{ 1 - \int_0^\infty\,dr_2 \,\frac{\partial v_\text{PS}(r_1,r_2;\varepsilon)}{\partial \varepsilon}\frac{\phi(r_2,\varepsilon)}{\phi(r_1,\varepsilon)} 
    - \int_0^\infty\,dr_2\,v_\text{PS}(r_1,r_2;\varepsilon)\left[\frac{1}{\phi(r_1,\varepsilon)}\frac{\partial \phi(r_2,\varepsilon)}{\partial\varepsilon} - \frac{\phi(r_2,\varepsilon)}{\phi^2(r_1,\varepsilon)}\frac{\partial\phi(r_1,\varepsilon)}{\partial\varepsilon}\right] \right\}.
\end{align}
We use the following identity, which is valid for any function $f(r,\varepsilon)$:
\begin{equation}
    \label{eq:useful_identity}
    \frac{d}{dr}\left[\phi^2(r,\varepsilon) f(r,\varepsilon)\right] = \phi^2(r,\varepsilon)\left[2\frac{\phi'(r,\varepsilon)}{\phi(r,\varepsilon)}f(r,    \varepsilon) + f'(r,\varepsilon)\right] = \phi^2(r,\varepsilon)\left[2 x(r,\varepsilon) f(r,\varepsilon) + f'(r,\varepsilon)\right],
\end{equation}
where in our case $f(r,\varepsilon) = \partial x(r,\varepsilon)/\partial\varepsilon$ and we can substitute Eq.~\eqref{eq:useful_identity} in the LHS of equation Eq.~\eqref{eq:divide_by_phi}.
\begin{align}
    \label{eq:substitute_useful_identity}
    &\frac{1}{\phi^2(r_1,\varepsilon)}\frac{d}{dr_1}\left[ \phi^2(r_1,\varepsilon) \frac{\partial x(r_1,\varepsilon)}{\partial\varepsilon}\right] =\\
\nonumber
    & = -2 \left\{ 1 - \int_0^\infty\,dr_2 \,\frac{\partial v_\text{PS}(r_1,r_2;\varepsilon)}{\partial \varepsilon}\frac{\phi(r_2,\varepsilon)}{\phi(r_1,\varepsilon)} 
    - \int_0^\infty\,dr_2\,v_\text{PS}(r_1,r_2;\varepsilon)\left[\frac{1}{\phi(r_1,\varepsilon)}\frac{\partial \phi(r_2,\varepsilon)}{\partial\varepsilon} - \frac{\phi(r_2,\varepsilon)}{\phi^2(r_1,\varepsilon)}\frac{\partial\phi(r_1,\varepsilon)}{\partial\varepsilon}\right] \right\}.
\end{align}
Then we can multiply both sides of Eq.~\eqref{eq:substitute_useful_identity} by $\phi^2(r_1,\varepsilon)$ and integrate between $r=0$ and $r=R_c$:
\begin{align}
    \label{eq:integrate_0_Rc}
    \phi^2(R_c,\varepsilon) \frac{\partial x(R_c,\varepsilon)}{\partial\varepsilon} &= -2\bigg\{ \int_0^{R_c}\,dr_1\, \phi^2(r_1) - \int_0^{R_c}\,dr_1\int_0^\infty\,dr_2\, \phi(r_1,\varepsilon)\frac{\partial v_\text{PS}}{\partial\varepsilon}(r_1,r_2;\varepsilon)\phi(r_2,\varepsilon) \\
\nonumber
    & - \int_0^{R_c}\,dr_1 \int_0^\infty\,dr_2\, v_\text{PS}(r_1,r_2;\varepsilon)\left[\phi(r_1,\varepsilon)\frac{\partial \phi(r_2,\varepsilon)}{\partial\varepsilon} - \phi(r_2,\varepsilon)\frac{\partial\phi(r_1,\varepsilon)}{\partial\varepsilon} \right]\bigg\}.
\end{align}
Further simplifications can be obtained assuming that 
\begin{enumerate}
    \item $v_\text{PS}$ is localized within $R_c$: $v_\text{PS}(r_1>R_c,r_2,\varepsilon)=v_\text{PS}(r_1,r_2>R_c,\varepsilon) = 0$.
    \item $v_\text{PS}$ is symmetric for change in the spatial coordinates: $v_\text{PS}(r_1,r_2;\varepsilon)=v_\text{PS}(r_2,r_1;\varepsilon)$.
\end{enumerate}
With these assumptions we achieve that the variables $r_1$ and $r_2$ live now in the same domain where double integrals involving $v_\text{PS}$ are defined and that the integral appearing in the second line of Eq.~\eqref{eq:integrate_0_Rc} vanishes, since it is given by the integration of the product of a symmetric function $(v_\text{PS}(r_1,r_2))$ for an antisymmetric one $\left(\phi(r_1)\frac{\partial \phi(r_2)}{\partial \varepsilon} - \phi(r_2) \frac{\partial\phi(r_1)}{\partial\varepsilon}\right)$.
The overall equation reduces to
\begin{equation}
    \label{eq:clean}
    \frac{\partial x(R_c,\varepsilon)}{\partial\varepsilon} = \frac{\partial}{\partial\varepsilon}\frac{d}{dr}\text{ln}\phi(r,\varepsilon)\bigg\rvert_{R_c} = - \frac{2}{\phi^2(R_c,\varepsilon)} \int_0^{R_c}\,dr_1\int_0^{R_c}\,dr_2\, \phi(r_1,\varepsilon) \left[\delta(r_1,r_2) - \frac{\partial v_\text{PS}(r_1,r_2;\varepsilon)}{\partial\varepsilon}\right] \phi(r_2,\varepsilon).
\end{equation}
Eq.~\eqref{eq:clean} can also be rewritten adopting a braket notation with
\begin{equation}
    \label{eq:clean_braket}
        \frac{\partial}{\partial\varepsilon}\frac{d}{dr}\text{ln}\phi(r,\varepsilon)\bigg\rvert_{R_c} = - \frac{2}{\phi^2(R_c,\varepsilon)} \bigg\langle \phi \bigg\rvert 1 - \frac{\partial v_\text{PS}(\varepsilon)}{\partial\varepsilon} \bigg\rvert\phi\bigg\rangle_{R_c},
\end{equation}
with $\braket{\phi}{\phi}_{R_c}$ meaning that the integration has been restricted within the core radius $R_c$. 
\section{Details of the linear system of equations for the embedding construction}
We delve into the details of the linear system of equations defined by the embedding construction:
\begin{equation}
    \label{eq:embedding_ham_supp}
    \begin{pmatrix}
        h_0  &   V  \\
        \overline{V}^\dagger & \sigma
    \end{pmatrix}
    \begin{pmatrix}
        \ket{\phi_i} \\
        \ket{q_i}
    \end{pmatrix}
    = \varepsilon_i
    \begin{pmatrix}
        \ket{\phi_i} \\
        \ket{q_i}
    \end{pmatrix},
\end{equation}
with $h_0 = T +v_\text{loc}$.
In the general case of a nonhermitian eigenvalue problem, right and left eigenvectors have to be considered.
However, within a pseudopotential generation scheme the PS orbitals $\ket{\phi_i}$ can always be taken as real valued, leaving the distinction of right and left eigenvectors to the $\ket{q_i}$ degrees of freedom.
Four equations can be derived as follows:
\begin{align}
   h_0 \ket{\phi_i} + V\ket{q_i^R} =& \varepsilon_i \ket{\phi_i}\\
   \overline{V}^\dagger\ket{\phi_i} +\sigma\ket{q_i^R} =&\varepsilon_i \ket{q_i^R}\\
   \bra{\phi_i}h_0 + \bra{q_i^L}\overline{V}^\dagger =&\varepsilon_i\bra{\phi_i}\\
   \bra{\phi_i}V + \bra{q_i^L}\sigma =& \varepsilon_i\bra{q_i^L}
\end{align}
All relevant quantities can be obtained from these 4 equalities.
In particular the localized projectors $\ket{\chi_i}$ are given as
\begin{align}
    \ket{\chi_i}&=(\varepsilon_i-h_0)\ket{\phi_i}=V\ket{q_i^R}\\
    \bra{\chi_i}&=\bra{\phi_i}(\varepsilon_i-h_0)=\bra{q_i^L}\overline{V}^\dagger,
\end{align}
and the $B$ matrix of the ultrasoft pseudopotential (USPP) formalism is obtained from
\begin{align}
    B_{ij} &= \braket{\phi_i}{\chi_j}=\bra{q_i^L}\varepsilon_i-\sigma\ket{q_j^R}\\
    B^\dagger_{ij} &= \braket{\chi_i}{\phi_j} = \bra{q_i^L}\varepsilon_j-\sigma\ket{q_j^R},
\end{align}
linking the Hermiticity condition of the $B_{ij}$ matrix to the augmentation charges $Q_{ij}=\braket{q_i}{q_j}$ as found for the USPP framework~\cite{VanderbiltSoft1990Supp}:
\begin{equation}
    \label{eq:hermiticity_B_supp}
    B_{ij}-B^\dagger_{ij} = Q_{ij}(\varepsilon_i-\varepsilon_j)
\end{equation}
Additionally, the matrix products of $h_0$ can be related to those of the bath $\sigma$ as
\begin{equation}
    \bra{\phi^L_i}\varepsilon_i - h_0\ket{\phi_j^R} = \bra{q^L_i}\varepsilon_i - \sigma \ket{q^R_j}.
\end{equation}
Finally, the matrix elements of the couplings $V$ and $\overline{V}^\dagger$ can be characterized by the overlap operator of the $\ket{\chi}$ projectors as
\begin{equation}
    \bra{q_i^L}\overline{V}^\dagger V\ket{q_j^R} = \braket{\chi_i}{\chi_j}
\end{equation}

\section{Variational construction of the basis set}
We show explicitly that the choice of our basis is consistent with a variational principle related to the minimization of a spread functional in the space spanned by the $\ket{\chi_i}$ functions, that we define as
\begin{equation}
    \Omega_\chi[b_i]=\sum_i\left[\braket{\chi_i}{\chi_i} - \sum_k|\braket{\chi_i}{b_k}|^2\right]+\sum_{i}\lambda_i(\braket{b_i}{b_i}-1),
\end{equation}
where Lagrange multipliers $\lambda_i$ are introduced to enforce the normalization of the basis states.
We then consider an expansion of the states $\ket{b_i}$ in terms of the $\ket{\chi_i}$ as
\begin{align}
    \ket{b_i} = \sum_j \ket{\chi_j}U_{ji}\\
    \bra{b_i} = \sum_j U^\dagger_{ij}\bra{\chi_j}
\end{align}
We then consider the variation of the functional with respect to $U^\dagger_{ij}$
\begin{equation}
    \frac{\delta\Omega_\chi}{\delta U^\dagger_{ij}} = -\sum_{k k'} S_{jk} S_{kk'}U_{k'i} + \sum_k S_{jk}U_{ki}\lambda_i,
\end{equation}
where we have introduced the overlap matrix $S_{ij}=\braket{\chi_i}{\chi_j}$.
The functional is made stationary by
\begin{equation}
    \sum_{k'}S_{kk'}U_{k'i} = U_{ki}\lambda_i,
\end{equation}
which is the eigenvalue equation for the overlap matrix with eigenvalue $-\lambda_i$.
The matrix $U_{ij}$ is then the matrix of the eigenvectors of $S_{ij}$.
The normalization condition of the basis states returns the following condition for the basis set:
\begin{equation}
    \ket{b_i} = \frac{1}{\sqrt{\lambda_i}}\sum_{j}\ket{\chi_j}U_{ji}.
\end{equation}
In a practical calculation one can restrict the number of basis states to $N_b< N$, where $N$ is the number of $\ket{\chi_i}$ projectors, truncating all states with eigenvalues $\lambda_i$ smaller than a given threshold.
The value of the functional when evaluated with a set of $N_b$ basis states returns
\begin{equation}
    \Omega[\{b_i\}]=\sum_{i=1}^{N}\lambda_i -\sum_{i=1}^{N_b}\lambda_i.
\end{equation}
In particular the value of the functional is zero for $N_b=N$ and equal to the sum of the eigenvalues of the truncated states otherwise, measuring the violation of completeness of the basis $\{\ket{b_i}\}$ in the space spanned by the $\ket{\chi_i}$ projectors.
%
%
%

\section{Numerical details for the generation of pseudopotentials}
We report the numerical details of the generation of SOP pseudopotentials in the case of the copper $d$ electrons and erbium $f$ electrons presented in the Letter.
The matrix elements of the screened pseudopotentials are also reported.
The uspp-736 code is used to generate the atomic orbitals and local potentials.
The Kohn-Sham exchange correlation functional used is PBE.
\subsection{Copper dynamical pseudopotential}
We generate the dynamical SOP pseudopotential for the copper (Cu) atom in the $d$ channel.
The reference atomic configuration is $3s^2\,3p^6\,3d^{9.5}\,4s^{1.5}$.
We consider a total of 7 reference energies, matching the logarithmic derivative of the AE and PS orbitals for energies up to 50 Ry.
The local potential and the PS orbitals are obtained through a polynomial pseudization at a cutoff radius $R_c=2.0$ (a.u.).
The reference energies considered in the pseudization are $\varepsilon^\text{ref}=[-0.5221,\,\, 5,\,\, 15\,\,, 22\,\,, 30,\,\, 40\,\,, 50]$ Ry.
We report the position of the poles and the matrix elements of the SOP pseudopotential in Tab.~\ref{tab:Cu_d}.
In the case of this pseudopotential, the chosen set of reference energies leads to appearence of pairs of complex conjugate poles, with complex conjugate residue matrices. 
Even with this combination the dynamical pseudopotential is a Hermitian operator when evaluated for real energies.
\subsection{Erbium dynamical pseudopotential}
We generate the dynamical SOP pseudopotential for the erbium (Er) atom in the $f$ channel.
The reference atomic configuration is the ionic $5s^2\,5p^6\,4f^{12}$ corresponding to Er$^{2+}$.
We consider a total of 7 reference energies, matching the logarithmic derivative of the AE and PS orbitals for energies up to $55$ Ry.
The local potential and the PS orbitals are obtained through a polynomial pseudization at a cutoff radius $R_c=1.5$ (a.u.).
The reference energies considered in the pseudization are $\varepsilon^\text{ref}=[-1.2816, \,\,5, \,\,15, \,\,25, \,\,35, \,\,45, \,\,55]$ Ry.
We report the position of the poles and the matrix elements of the screened residues of the SOP pseudopotential in Tab.~\ref{tab:Er_f}.
For the chosen set of reference energies the dynamical pseudopotential has real poles and Hermitian residues.
\begin{table}[]
    \centering
    \caption{Poles [Ry] and (screened) residues [Ry$^2$] of the Cu pseudopotential in the $d$-channel.}
    \begin{tabular}{c|c c c}
    \hline
    \hline 
    Pole &  \multicolumn{3}{c}{Res. (screened)} \\
    \hline 
113.6926 - 0.0000 i & -0.3411 - 0.0000 i & -3.5386 - 0.0000 i & -36.2824 - 0.0000 i\\
 & -3.5386 - 0.0000 i & -36.7113 - 0.0000 i & -376.4157 - 0.0000 i\\
 & -36.2824 - 0.0000 i & -376.4157 - 0.0000 i & -3859.5360 - 0.0000 i\\
\hline
48.0927 - 10.1171 i & -0.2073 + 0.1220 i & -4.7286 + 10.7031 i & -3.6076 - 1.9587 i\\
 & -4.7286 + 10.7031 i & 116.7979 + 557.0265 i & -198.1007 + 24.9515 i\\
 & -3.6076 - 1.9587 i & -198.1007 + 24.9515 i & -3.0741 - 69.9794 i\\
\hline
48.0927 + 10.1171 i & -0.2073 - 0.1220 i & -4.7286 - 10.7031 i & -3.6076 + 1.9587 i\\
 & -4.7286 - 10.7031 i & 116.7979 - 557.0265 i & -198.1007 - 24.9515 i\\
 & -3.6076 + 1.9587 i & -198.1007 - 24.9515 i & -3.0741 + 69.9794 i\\
\hline
30.8418 - 0.0000 i & 0.0520 - 0.0000 i & -0.8688 - 0.0000 i & 1.5387 - 0.0000 i\\
 & -0.8688 - 0.0000 i & 14.5235 - 0.0000 i & -25.7219 - 0.0000 i\\
 & 1.5387 - 0.0000 i & -25.7219 - 0.0000 i & 45.5551 - 0.0000 i\\
\hline
5.2131 - 4.0387 i & -0.0301 + 0.0096 i & -3.1809 + 2.5336 i & -2.7653 + 0.6661 i\\
 & -3.1809 + 2.5336 i & -267.1540 + 450.2113 i & -302.4132 + 206.4567 i\\
 & -2.7653 + 0.6661 i & -302.4132 + 206.4567 i & -252.7281 + 41.5094 i\\
\hline
5.2131 + 4.0387 i & -0.0301 - 0.0096 i & -3.1809 - 2.5336 i & -2.7653 - 0.6661 i\\
 & -3.1809 - 2.5336 i & -267.1540 - 450.2113 i & -302.4132 - 206.4567 i\\
 & -2.7653 - 0.6661 i & -302.4132 - 206.4567 i & -252.7281 - 41.5094 i\\
\hline
1.2442 - 0.0000 i & 0.0153 - 0.0000 i & 1.5309 - 0.0000 i & 1.5023 - 0.0000 i\\
 & 1.5309 - 0.0000 i & 153.0327 - 0.0000 i & 150.1694 - 0.0000 i\\
 & 1.5023 - 0.0000 i & 150.1694 - 0.0000 i & 147.3596 - 0.0000 i\\
    \hline
    \end{tabular}
    \label{tab:Cu_d}
\end{table}
\begin{table}[]
    \centering
    \caption{Poles [Ry] and (screened) residues [Ry$^2$] of the Er pseudopotential in the $f$-channel.}
    \begin{tabular}{c|c c c}
    \hline
    \hline 
    Pole &  \multicolumn{3}{c}{Res. (screened)} \\
    \hline 
214.5115 - 0.0000 i & -174.3217 - 0.0000 i & -1231.4795 - 0.0000 i & -1048.3375 - 0.0000 i\\
 & -1231.4795 - 0.0000 i & -8699.6745 - 0.0000 i & -7405.8848 - 0.0000 i\\
 & -1048.3375 - 0.0000 i & -7405.8848 - 0.0000 i & -6304.5037 - 0.0000 i\\
\hline
-79.0617 - 0.0000 i & -56.0337 - 0.0000 i & -594.9246 - 0.0000 i & -306.6263 - 0.0000 i\\
 & -594.9246 - 0.0000 i & -6316.4742 - 0.0000 i & -3255.5332 - 0.0000 i\\
 & -306.6263 - 0.0000 i & -3255.5332 - 0.0000 i & -1677.9134 - 0.0000 i\\
\hline
-0.1944 - 0.0000 i & 0.2210 - 0.0000 i & 1.3588 - 0.0000 i & -1.5372 - 0.0000 i\\
 & 1.3588 - 0.0000 i & 8.3553 - 0.0000 i & -9.4526 - 0.0000 i\\
 & -1.5372 - 0.0000 i & -9.4526 - 0.0000 i & 10.6941 - 0.0000 i\\
\hline
51.9961 - 0.0000 i & 17.1474 - 0.0000 i & -43.2236 - 0.0000 i & 56.4740 - 0.0000 i\\
 & -43.2236 - 0.0000 i & 108.9543 - 0.0000 i & -142.3545 - 0.0000 i\\
 & 56.4740 - 0.0000 i & -142.3545 - 0.0000 i & 185.9938 - 0.0000 i\\
\hline
20.0510 - 0.0000 i & 0.2868 - 0.0000 i & 2.8518 - 0.0000 i & -5.4667 - 0.0000 i\\
 & 2.8518 - 0.0000 i & 28.3605 - 0.0000 i & -54.3639 - 0.0000 i\\
 & -5.4667 - 0.0000 i & -54.3639 - 0.0000 i & 104.2094 - 0.0000 i\\
\hline
25.8854 - 0.0000 i & -1.7477 - 0.0000 i & 9.3699 - 0.0000 i & 1.3702 - 0.0000 i\\
 & 9.3699 - 0.0000 i & -50.2352 - 0.0000 i & -7.3459 - 0.0000 i\\
 & 1.3702 - 0.0000 i & -7.3459 - 0.0000 i & -1.0742 - 0.0000 i\\
\hline
41.1268 - 0.0000 i & -2.2470 - 0.0000 i & -15.5076 - 0.0000 i & 6.9548 - 0.0000 i\\
 & -15.5076 - 0.0000 i & -107.0239 - 0.0000 i & 47.9981 - 0.0000 i\\
 & 6.9548 - 0.0000 i & 47.9981 - 0.0000 i & -21.5262 - 0.0000 i\\
\hline
    \end{tabular}
    \label{tab:Er_f}
\end{table}
%
%
\section{Details of the screening and descreening procedure}
The ionic (descreened) dynamical pseudopotential is obtained from the screened one through
\begin{equation}
    v_\text{PS}^\text{ion}(\omega)
    =
    v_\text{PS}(\omega)
    +
    \int dr\, v_\text{loc}(r)\, f(r,\omega)\,
    \frac{\partial v_\text{PS}}{\partial \omega},
\end{equation}
where
\begin{equation}
v_\text{loc}(r)=v_\text{Hxc}(r)+v_\text{loc}^\text{ion}(r),
\end{equation}
and \(f(r,\omega)\) is a normalized function,
\begin{equation}
\int_0^{R_c} dr\, f(r,\omega)=1 \qquad \forall\,\omega,
\end{equation}
introduced to reconstruct the spatial profile of the augmentation charge inside the core region.
Within the SOP representation, the dynamical pseudopotential is written as
\begin{equation}
    v_\text{PS}(\omega)
    =
    \sum_{\alpha\beta} \ket{b_\alpha}
    \sum_s \frac{\Gamma^s_{\alpha\beta}}{\omega-\Omega_s}
    \bra{b_\beta},
\end{equation}
where \(\Gamma^s_{\alpha\beta}\) are the matrix-valued residues and \(\Omega_s\) are the poles.
In this representation, screening and descreening can be carried out by modifying the residues while leaving the poles unchanged.
To do so, one evaluates the potential at a set of energies \(\{\varepsilon_i\}\), whose number is chosen to match the number of poles, for instance by taking the same reference energies used in the construction of the pseudopotential.
For a fixed matrix element \((\alpha,\beta)\), the screening/descreening relation becomes
\begin{equation}
\label{eq:screening_sop}
    \sum_s \frac{1}{\varepsilon_i-\Omega_s}\,\widetilde{\Gamma}^s_{\alpha\beta}
    =
    \sum_s \frac{1}{\varepsilon_i-\Omega_s}
    \left(
    1-\frac{\overline f_i}{\varepsilon_i-\Omega_s}
    \right)
    \Gamma^s_{\alpha\beta},
\end{equation}
where \(\widetilde{\Gamma}^s_{\alpha\beta}\) are the residues of the ionic potential, and
\begin{equation}
    \overline f_i
    =
    \int_0^{R_c} dr\, v_\text{loc}(r)\, f(r,\varepsilon_i).
\end{equation}
Equation~\eqref{eq:screening_sop} can be written more compactly by introducing the matrices
\begin{align}
    A^\text{ion}_{is} &= \frac{1}{\varepsilon_i-\Omega_s},\\
    A^\text{scr}_{is} &= \frac{1}{\varepsilon_i-\Omega_s}
    \left(
    1-\frac{\overline f_i}{\varepsilon_i-\Omega_s}
    \right),
\end{align}
so that
\begin{equation}
    \sum_s A^\text{ion}_{is}\,\widetilde{\Gamma}^s_{\alpha\beta}
    =
    \sum_s A^\text{scr}_{is}\,\Gamma^s_{\alpha\beta}.
\end{equation}
The residues of the screened and ionic potentials are then obtained from one another through
\begin{align}
    \widetilde{\Gamma}^s_{\alpha\beta}
    &=
    \sum_{im}
    \left(A^\text{ion}\right)^{-1}_{si}
    A^\text{scr}_{im}\,
    \Gamma^m_{\alpha\beta},\\
    \Gamma^s_{\alpha\beta}
    &=
    \sum_{im}
    \left(A^\text{scr}\right)^{-1}_{si}
    A^\text{ion}_{im}\,
    \widetilde{\Gamma}^m_{\alpha\beta}.
\end{align}
\section{Interpolation properties}
\begin{figure*}[t]
    \centering
    \includegraphics[width=0.49\textwidth]{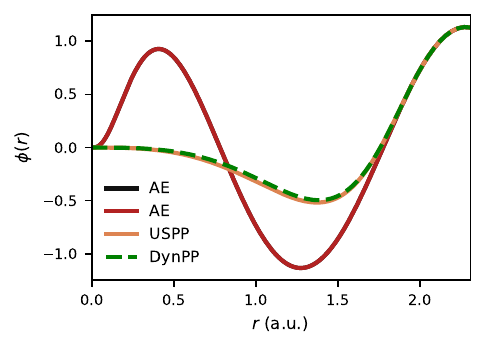}
    \hfill
    \includegraphics[width=0.49\textwidth]{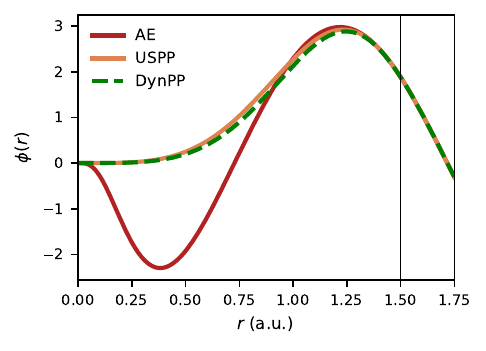}
    \caption{
    Reconstruction of pseudo-orbitals at energies not included among the reference states used in the generation procedure.
    The all-electron (AE) wavefunction is computed for the unbound scattering state at $10$ Ry for the $d$ channel of copper (left) and the $f$ channel of erbium (right).
    The corresponding dynamical-pseudopotential (DynPP) wavefunction is reconstructed at the same scattering energy and compared with an ultrasoft pseudo-orbital generated directly at $10$ Ry through polynomial pseudization of the AE wavefunction. The core radius is signaled with a vertical black line.
    }
    \label{fig:wfc_10ry_interp}
\end{figure*}

We assess the extent to which the dynamical pseudopotential reconstructs pseudo-orbitals at energies that are not included among the reference states used in the generation procedure.
To this end, we consider the unbound scattering state at $10$ Ry for the copper dynamical pseudopotential in the $d$ channel and for the erbium dynamical pseudopotential in the $f$ channel.
As a benchmark for the quality of the reconstruction, we also generate, at the same energy, an ultrasoft pseudo-orbital using the same pseudization algorithm and the same set of parameters used in the generation procedure for the reference states.
The results are shown in Fig.~\ref{fig:wfc_10ry_interp}.
In both cases, the dynamical pseudopotential reconstructs the pseudo-orbital with excellent accuracy even at an energy that was not used in the pseudization procedure, demonstrating very good interpolation properties.
The resulting pseudo-orbitals are nearly indistinguishable from those obtained by carrying out an explicit polynomial pseudization of the all-electron state at the same energy.
\section{Smoothness of the pseudo-orbitals}
\begin{figure}
    \centering
    \includegraphics[width=0.8\linewidth]{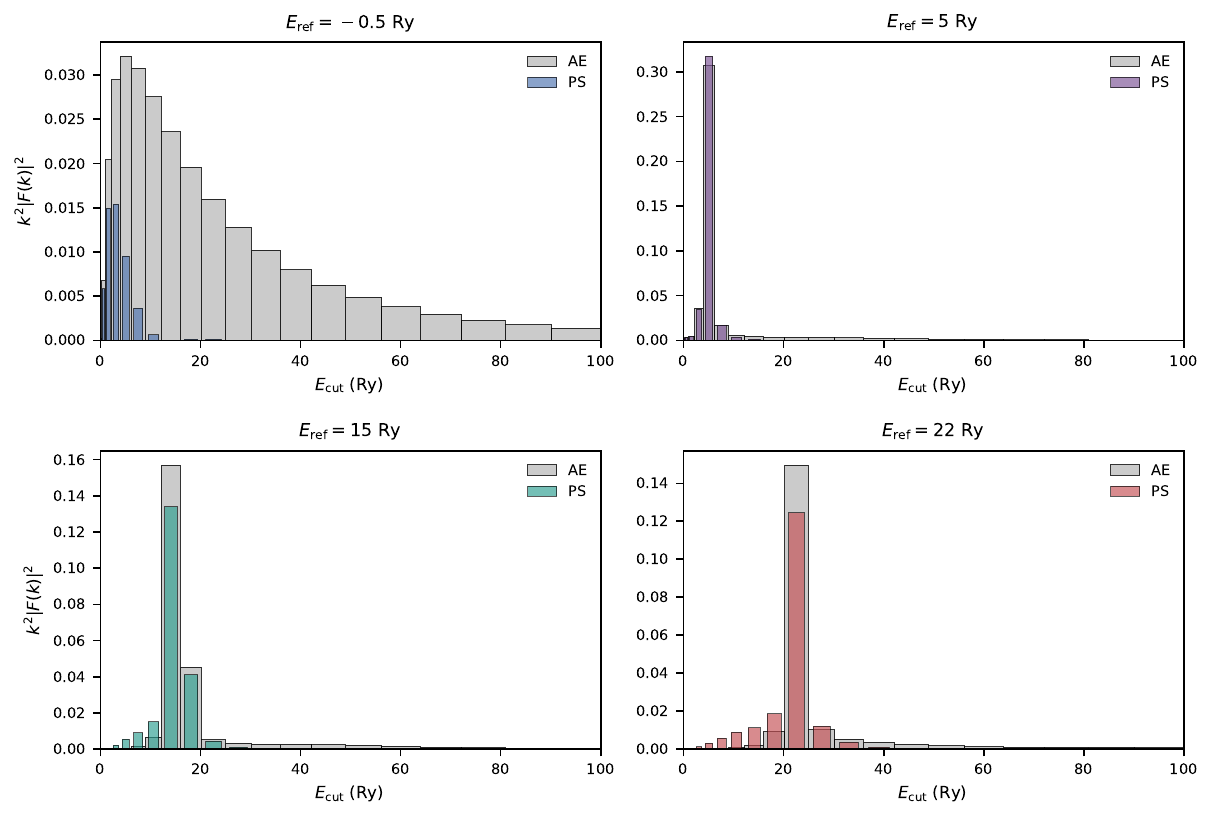}
    \caption{Radial momentum distribution of the pseudo-orbitals of the copper (Cu) dynamical pseudopotential for selected reference energies used in the generation.
    The state at $E_{\mathrm{ref}}=-0.5$ Ry is the bound Cu $3d$ state, while the other states are unbound.
    For unbound states, the transform is peaked around the wavevector corresponding to a free spherical wave of the same energy.}
    \label{fig:soft_histo_cu}
\end{figure}

\begin{figure}
    \centering
    \includegraphics[width=0.8\linewidth]{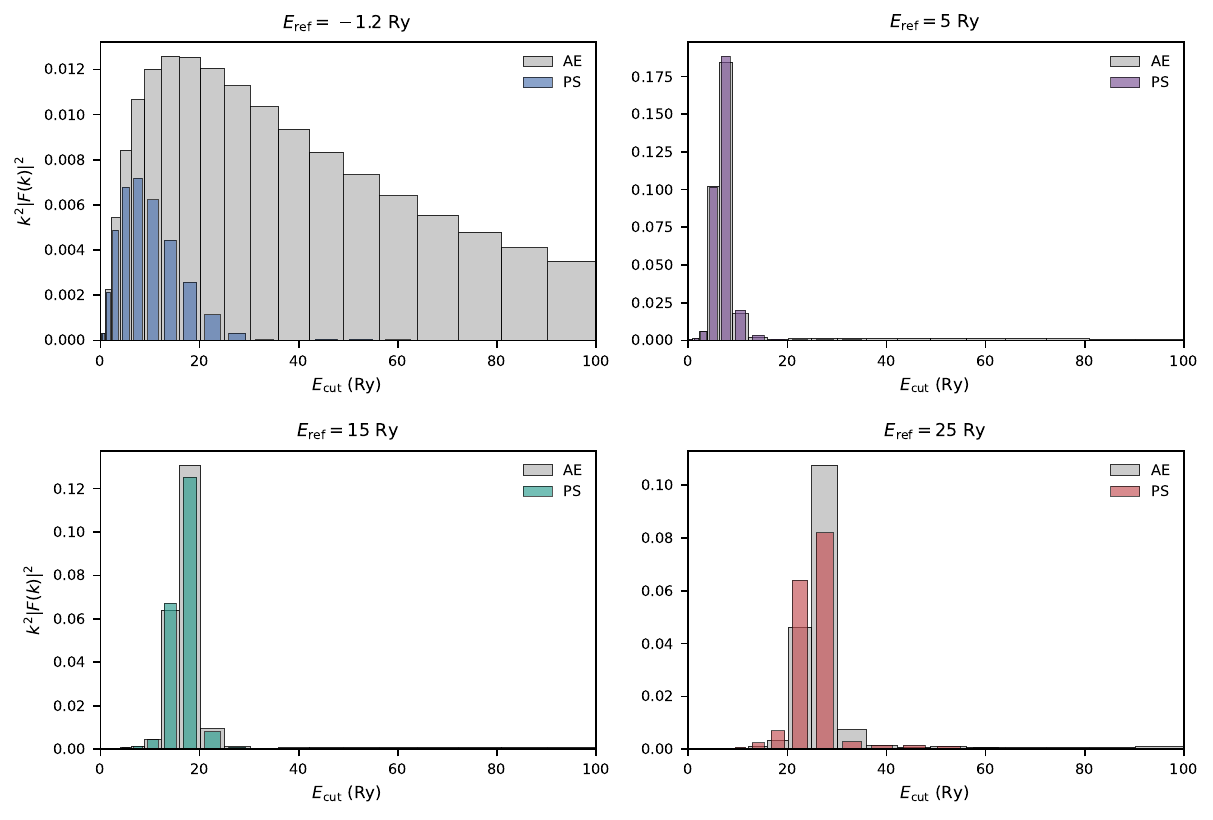}
    \caption{Radial momentum distribution of the pseudo-orbitals of the erbium (Er) dynamical pseudopotential for selected reference energies used in the generation.
    The state at $E_{\mathrm{ref}}=-1.2$ Ry is the bound Er $4f$ state, while the other states are unbound.
    For unbound states, the transform is peaked around the wavevector corresponding to a free spherical wave of the same energy.}
    \label{fig:soft_histo_er}
\end{figure}

We assess the smoothness of the pseudo-orbitals by estimating the largest momentum required for an accurate representation of the orbital within a plane-wave expansion.
To this end, we consider the spherical Bessel transform of the radial wavefunction.
For a given angular-momentum channel $l$, we write the wavefunction at a given energy as
\begin{equation}
\psi_{lm}(\mathbf r)=R_l(r)Y_{lm}(\theta,\phi),
\end{equation}
where $Y_{lm}$ is a spherical harmonic and $m$ labels the eigenvalue of the angular-momentum operator along the $z$ direction.
Introducing the reduced radial wavefunction $u_l(r)=rR_l(r)$, the corresponding radial transform is proportional to
\begin{equation}
    \tilde{u}_l(k) \propto \int_0^{\infty} dr\, r\,u_l(r)\, j_l(kr),
\end{equation}
where $j_l(kr)$ is the spherical Bessel function of order $l$ and the normalization factor is omitted, since it is the same for both AE and PS wavefunctions.
In three-dimensional momentum space, the contribution of a shell of radius $k$ is weighted by its phase-space factor.
The corresponding radial momentum distribution is therefore proportional to
\begin{equation}
w(k)=k^2 |\tilde{u}_l(k)|^2,
\end{equation}
where the factor $k^2$ arises from the integration over the spherical shell in momentum space.
In Fig.~\ref{fig:soft_histo_cu} we report this distribution for the pseudo-orbitals of the copper dynamical pseudopotential in the $d$ channel, plotted as a function of the energy (in Ry) of a free plane wave with the same wavevector.
Since distances are measured in Bohr, this energy is related to the wavevector by $E=k^2$.
The same analysis is reported in Fig.~\ref{fig:soft_histo_er} for the $f$ channel of the erbium dynamical pseudopotential.
In both cases, we compare the pseudo-orbital distributions with those of the corresponding all-electron states generated at the same energy.
For bound states, the momentum cutoff is reduced by approximately one order of magnitude: in copper, from about $100$ Ry for the AE orbital to about $10$ Ry for the PS orbital, and in erbium from about $300$ Ry to about $30$ Ry.
For unbound states with energy $\varepsilon>0$, the transform is dominated by the long-range tail of the wavefunction, which asymptotically behaves as a free spherical wave with wavevector satisfying $k^2=\varepsilon$.
As a consequence, the momentum distributions of AE and PS orbitals are similar at large scales, although the PS orbitals show an enhanced redistribution of weight toward smaller wavevectors.
Accordingly, an accurate representation of an unbound pseudo-orbital generated at energy $\varepsilon>0$ still requires a plane-wave cutoff of the same order as the energy eigenvalue itself.


\begin{thebibliography}{50}%
\makeatletter
\providecommand \@ifxundefined [1]{%
 \@ifx{#1\undefined}
}%
\providecommand \@ifnum [1]{%
 \ifnum #1\expandafter \@firstoftwo
 \else \expandafter \@secondoftwo
 \fi
}%
\providecommand \@ifx [1]{%
 \ifx #1\expandafter \@firstoftwo
 \else \expandafter \@secondoftwo
 \fi
}%
\providecommand \natexlab [1]{#1}%
\providecommand \enquote  [1]{``#1''}%
\providecommand \bibnamefont  [1]{#1}%
\providecommand \bibfnamefont [1]{#1}%
\providecommand \citenamefont [1]{#1}%
\providecommand \href@noop [0]{\@secondoftwo}%
\providecommand \href [0]{\begingroup \@sanitize@url \@href}%
\providecommand \@href[1]{\@@startlink{#1}\@@href}%
\providecommand \@@href[1]{\endgroup#1\@@endlink}%
\providecommand \@sanitize@url [0]{\catcode `\\12\catcode `\$12\catcode `\&12\catcode `\#12\catcode `\^12\catcode `\_12\catcode `\%12\relax}%
\providecommand \@@startlink[1]{}%
\providecommand \@@endlink[0]{}%
\providecommand \url  [0]{\begingroup\@sanitize@url \@url }%
\providecommand \@url [1]{\endgroup\@href {#1}{\urlprefix }}%
\providecommand \urlprefix  [0]{URL }%
\providecommand \Eprint [0]{\href }%
\providecommand \doibase [0]{http://dx.doi.org/}%
\providecommand \selectlanguage [0]{\@gobble}%
\providecommand \bibinfo  [0]{\@secondoftwo}%
\providecommand \bibfield  [0]{\@secondoftwo}%
\providecommand \translation [1]{[#1]}%
\providecommand \BibitemOpen [0]{}%
\providecommand \bibitemStop [0]{}%
\providecommand \bibitemNoStop [0]{.\EOS\space}%
\providecommand \EOS [0]{\spacefactor3000\relax}%
\providecommand \BibitemShut  [1]{\csname bibitem#1\endcsname}%
\let\auto@bib@innerbib\@empty
\bibitem [{\citenamefont {Phillips}\ and\ \citenamefont {Kleinman}(1959)}]{PhillipsNew1959}%
  \BibitemOpen
  \bibfield  {author} {\bibinfo {author} {\bibfnamefont {J.~C.}\ \bibnamefont {Phillips}}\ and\ \bibinfo {author} {\bibfnamefont {L.}~\bibnamefont {Kleinman}},\ }\href {\doibase 10.1103/PhysRev.116.287} {\bibfield  {journal} {\bibinfo  {journal} {Phys. Rev.}\ }\textbf {\bibinfo {volume} {116}},\ \bibinfo {pages} {287} (\bibinfo {year} {1959})}\BibitemShut {NoStop}%
\bibitem [{\citenamefont {Heine}(1970)}]{HeineThepseudopotential1970}%
  \BibitemOpen
  \bibfield  {author} {\bibinfo {author} {\bibfnamefont {V.}~\bibnamefont {Heine}},\ }\href {\doibase https://doi.org/10.1016/S0081-1947(08)60069-7} {\emph {\bibinfo {title} {The Pseudopotential Concept}}},\ edited by\ \bibinfo {editor} {\bibfnamefont {H.}~\bibnamefont {Ehrenreich}}, \bibinfo {editor} {\bibfnamefont {F.}~\bibnamefont {Seitz}}, \ and\ \bibinfo {editor} {\bibfnamefont {D.}~\bibnamefont {Turnbull}},\ \bibinfo {series} {Solid State Physics}, Vol.~\bibinfo {volume} {24}\ (\bibinfo  {publisher} {Academic Press},\ \bibinfo {year} {1970})\ pp.\ \bibinfo {pages} {1--36}\BibitemShut {NoStop}%
\bibitem [{\citenamefont {Hamann}\ \emph {et~al.}(1979)\citenamefont {Hamann}, \citenamefont {Schl\"uter},\ and\ \citenamefont {Chiang}}]{HamannNormconserving1979}%
  \BibitemOpen
  \bibfield  {author} {\bibinfo {author} {\bibfnamefont {D.~R.}\ \bibnamefont {Hamann}}, \bibinfo {author} {\bibfnamefont {M.}~\bibnamefont {Schl\"uter}}, \ and\ \bibinfo {author} {\bibfnamefont {C.}~\bibnamefont {Chiang}},\ }\href {\doibase 10.1103/PhysRevLett.43.1494} {\bibfield  {journal} {\bibinfo  {journal} {Phys. Rev. Lett.}\ }\textbf {\bibinfo {volume} {43}},\ \bibinfo {pages} {1494} (\bibinfo {year} {1979})}\BibitemShut {NoStop}%
\bibitem [{\citenamefont {Bachelet}\ \emph {et~al.}(1982)\citenamefont {Bachelet}, \citenamefont {Hamann},\ and\ \citenamefont {Schl\"uter}}]{BacheletPseudopotentials1982}%
  \BibitemOpen
  \bibfield  {author} {\bibinfo {author} {\bibfnamefont {G.~B.}\ \bibnamefont {Bachelet}}, \bibinfo {author} {\bibfnamefont {D.~R.}\ \bibnamefont {Hamann}}, \ and\ \bibinfo {author} {\bibfnamefont {M.}~\bibnamefont {Schl\"uter}},\ }\href {\doibase 10.1103/PhysRevB.26.4199} {\bibfield  {journal} {\bibinfo  {journal} {Phys. Rev. B}\ }\textbf {\bibinfo {volume} {26}},\ \bibinfo {pages} {4199} (\bibinfo {year} {1982})}\BibitemShut {NoStop}%
\bibitem [{\citenamefont {Kleinman}\ and\ \citenamefont {Bylander}(1982)}]{KleinmanEfficacious1982}%
  \BibitemOpen
  \bibfield  {author} {\bibinfo {author} {\bibfnamefont {L.}~\bibnamefont {Kleinman}}\ and\ \bibinfo {author} {\bibfnamefont {D.~M.}\ \bibnamefont {Bylander}},\ }\href {\doibase 10.1103/PhysRevLett.48.1425} {\bibfield  {journal} {\bibinfo  {journal} {Phys. Rev. Lett.}\ }\textbf {\bibinfo {volume} {48}},\ \bibinfo {pages} {1425} (\bibinfo {year} {1982})}\BibitemShut {NoStop}%
\bibitem [{\citenamefont {Vanderbilt}(1985)}]{Vanderbilt1985Optimally}%
  \BibitemOpen
  \bibfield  {author} {\bibinfo {author} {\bibfnamefont {D.}~\bibnamefont {Vanderbilt}},\ }\href {\doibase 10.1103/PhysRevB.32.8412} {\bibfield  {journal} {\bibinfo  {journal} {Phys. Rev. B}\ }\textbf {\bibinfo {volume} {32}},\ \bibinfo {pages} {8412} (\bibinfo {year} {1985})}\BibitemShut {NoStop}%
\bibitem [{\citenamefont {Troullier}\ and\ \citenamefont {Martins}(1991)}]{TroullerEfficient1990}%
  \BibitemOpen
  \bibfield  {author} {\bibinfo {author} {\bibfnamefont {N.}~\bibnamefont {Troullier}}\ and\ \bibinfo {author} {\bibfnamefont {J.~L.}\ \bibnamefont {Martins}},\ }\href {\doibase 10.1103/PhysRevB.43.1993} {\bibfield  {journal} {\bibinfo  {journal} {Phys. Rev. B}\ }\textbf {\bibinfo {volume} {43}},\ \bibinfo {pages} {1993} (\bibinfo {year} {1991})}\BibitemShut {NoStop}%
\bibitem [{\citenamefont {Vanderbilt}(1990)}]{VanderbiltSoft1990}%
  \BibitemOpen
  \bibfield  {author} {\bibinfo {author} {\bibfnamefont {D.}~\bibnamefont {Vanderbilt}},\ }\href {\doibase 10.1103/PhysRevB.41.7892} {\bibfield  {journal} {\bibinfo  {journal} {Phys. Rev. B}\ }\textbf {\bibinfo {volume} {41}},\ \bibinfo {pages} {7892} (\bibinfo {year} {1990})}\BibitemShut {NoStop}%
\bibitem [{\citenamefont {Rappe}\ \emph {et~al.}(1990)\citenamefont {Rappe}, \citenamefont {Rabe}, \citenamefont {Kaxiras},\ and\ \citenamefont {Joannopoulos}}]{Rappe1990Optimized}%
  \BibitemOpen
  \bibfield  {author} {\bibinfo {author} {\bibfnamefont {A.~M.}\ \bibnamefont {Rappe}}, \bibinfo {author} {\bibfnamefont {K.~M.}\ \bibnamefont {Rabe}}, \bibinfo {author} {\bibfnamefont {E.}~\bibnamefont {Kaxiras}}, \ and\ \bibinfo {author} {\bibfnamefont {J.~D.}\ \bibnamefont {Joannopoulos}},\ }\href {\doibase 10.1103/PhysRevB.41.1227} {\bibfield  {journal} {\bibinfo  {journal} {Phys. Rev. B}\ }\textbf {\bibinfo {volume} {41}},\ \bibinfo {pages} {1227} (\bibinfo {year} {1990})}\BibitemShut {NoStop}%
\bibitem [{\citenamefont {Hamann}(2013)}]{HamannOptimized2013}%
  \BibitemOpen
  \bibfield  {author} {\bibinfo {author} {\bibfnamefont {D.~R.}\ \bibnamefont {Hamann}},\ }\href {\doibase 10.1103/PhysRevB.88.085117} {\bibfield  {journal} {\bibinfo  {journal} {Phys. Rev. B}\ }\textbf {\bibinfo {volume} {88}},\ \bibinfo {pages} {085117} (\bibinfo {year} {2013})}\BibitemShut {NoStop}%
\bibitem [{\citenamefont {Hohenberg}\ and\ \citenamefont {Kohn}(1964)}]{HohenbergInhomogeneous1964}%
  \BibitemOpen
  \bibfield  {author} {\bibinfo {author} {\bibfnamefont {P.}~\bibnamefont {Hohenberg}}\ and\ \bibinfo {author} {\bibfnamefont {W.}~\bibnamefont {Kohn}},\ }\href {\doibase 10.1103/PhysRev.136.B864} {\bibfield  {journal} {\bibinfo  {journal} {Phys. Rev.}\ }\textbf {\bibinfo {volume} {136}},\ \bibinfo {pages} {B864} (\bibinfo {year} {1964})}\BibitemShut {NoStop}%
\bibitem [{\citenamefont {Kohn}\ and\ \citenamefont {Sham}(1965)}]{KohnSelfconsistent1965}%
  \BibitemOpen
  \bibfield  {author} {\bibinfo {author} {\bibfnamefont {W.}~\bibnamefont {Kohn}}\ and\ \bibinfo {author} {\bibfnamefont {L.~J.}\ \bibnamefont {Sham}},\ }\href {\doibase 10.1103/PhysRev.140.A1133} {\bibfield  {journal} {\bibinfo  {journal} {Phys. Rev.}\ }\textbf {\bibinfo {volume} {140}},\ \bibinfo {pages} {A1133} (\bibinfo {year} {1965})}\BibitemShut {NoStop}%
\bibitem [{\citenamefont {Marzari}\ \emph {et~al.}(2021)\citenamefont {Marzari}, \citenamefont {Ferretti},\ and\ \citenamefont {Wolverton}}]{MarzariElectronicstructure2021}%
  \BibitemOpen
  \bibfield  {author} {\bibinfo {author} {\bibfnamefont {N.}~\bibnamefont {Marzari}}, \bibinfo {author} {\bibfnamefont {A.}~\bibnamefont {Ferretti}}, \ and\ \bibinfo {author} {\bibfnamefont {C.}~\bibnamefont {Wolverton}},\ }\href {\doibase 10.1038/s41563-021-01013-3} {\bibfield  {journal} {\bibinfo  {journal} {Nature Materials}\ }\textbf {\bibinfo {volume} {20}},\ \bibinfo {pages} {736} (\bibinfo {year} {2021})}\BibitemShut {NoStop}%
\bibitem [{\citenamefont {Bosoni}\ \emph {et~al.}(2024)\citenamefont {Bosoni}, \citenamefont {Beal}, \citenamefont {Bercx}, \citenamefont {Blaha}, \citenamefont {Bl\"ugel}, \citenamefont {Br\"oder}, \citenamefont {Callsen}, \citenamefont {Cottenier}, \citenamefont {Degomme}, \citenamefont {Dikan}, \citenamefont {Eimre}, \citenamefont {Flage-Larsen}, \citenamefont {Fornari}, \citenamefont {Garcia}, \citenamefont {Genovese}, \citenamefont {Giantomassi}, \citenamefont {Huber}, \citenamefont {Janssen}, \citenamefont {Kastlunger}, \citenamefont {Krack}, \citenamefont {Kresse}, \citenamefont {K\"unhe}, \citenamefont {Lejaeghere}, \citenamefont {Madsen}, \citenamefont {Marzari}, \citenamefont {Michalicek}, \citenamefont {Mirhosseini}, \citenamefont {Müller}, \citenamefont {Petretto}, \citenamefont {Pickard}, \citenamefont {Poncé}, \citenamefont {Rignanese}, \citenamefont {Rubel}, \citenamefont {Ruh}, \citenamefont {Sluydts}, \citenamefont {Vanpoucke}, \citenamefont {Vijay}, \citenamefont {Wolloch}, \citenamefont
  {Wortmann}, \citenamefont {Yakutovich}, \citenamefont {Yu}, \citenamefont {Zadoks}, \citenamefont {Zhu},\ and\ \citenamefont {Pizzi}}]{BosoniDensityfunctional2024}%
  \BibitemOpen
  \bibfield  {author} {\bibinfo {author} {\bibfnamefont {E.}~\bibnamefont {Bosoni}}, \bibinfo {author} {\bibfnamefont {L.}~\bibnamefont {Beal}}, \bibinfo {author} {\bibfnamefont {M.}~\bibnamefont {Bercx}}, \bibinfo {author} {\bibfnamefont {P.}~\bibnamefont {Blaha}}, \bibinfo {author} {\bibfnamefont {S.}~\bibnamefont {Bl\"ugel}}, \bibinfo {author} {\bibfnamefont {J.}~\bibnamefont {Br\"oder}}, \bibinfo {author} {\bibfnamefont {M.}~\bibnamefont {Callsen}}, \bibinfo {author} {\bibfnamefont {S.}~\bibnamefont {Cottenier}}, \bibinfo {author} {\bibfnamefont {A.}~\bibnamefont {Degomme}}, \bibinfo {author} {\bibfnamefont {V.}~\bibnamefont {Dikan}}, \bibinfo {author} {\bibfnamefont {K.}~\bibnamefont {Eimre}}, \bibinfo {author} {\bibfnamefont {E.}~\bibnamefont {Flage-Larsen}}, \bibinfo {author} {\bibfnamefont {M.}~\bibnamefont {Fornari}}, \bibinfo {author} {\bibfnamefont {A.}~\bibnamefont {Garcia}}, \bibinfo {author} {\bibfnamefont {L.}~\bibnamefont {Genovese}}, \bibinfo {author} {\bibfnamefont {M.}~\bibnamefont
  {Giantomassi}}, \bibinfo {author} {\bibfnamefont {S.~P.}\ \bibnamefont {Huber}}, \bibinfo {author} {\bibfnamefont {H.}~\bibnamefont {Janssen}}, \bibinfo {author} {\bibfnamefont {G.}~\bibnamefont {Kastlunger}}, \bibinfo {author} {\bibfnamefont {M.}~\bibnamefont {Krack}}, \bibinfo {author} {\bibfnamefont {G.}~\bibnamefont {Kresse}}, \bibinfo {author} {\bibfnamefont {T.~D.}\ \bibnamefont {K\"unhe}}, \bibinfo {author} {\bibfnamefont {K.}~\bibnamefont {Lejaeghere}}, \bibinfo {author} {\bibfnamefont {M.}~\bibnamefont {Madsen}, \bibfnamefont {Georg K. H.and~Marsman}}, \bibinfo {author} {\bibfnamefont {N.}~\bibnamefont {Marzari}}, \bibinfo {author} {\bibfnamefont {G.}~\bibnamefont {Michalicek}}, \bibinfo {author} {\bibfnamefont {H.}~\bibnamefont {Mirhosseini}}, \bibinfo {author} {\bibfnamefont {T.~M.~A.}\ \bibnamefont {Müller}}, \bibinfo {author} {\bibfnamefont {G.}~\bibnamefont {Petretto}}, \bibinfo {author} {\bibfnamefont {C.~J.}\ \bibnamefont {Pickard}}, \bibinfo {author} {\bibfnamefont {S.}~\bibnamefont
  {Poncé}}, \bibinfo {author} {\bibfnamefont {G.-M.}\ \bibnamefont {Rignanese}}, \bibinfo {author} {\bibfnamefont {O.}~\bibnamefont {Rubel}}, \bibinfo {author} {\bibfnamefont {T.}~\bibnamefont {Ruh}}, \bibinfo {author} {\bibfnamefont {M.}~\bibnamefont {Sluydts}}, \bibinfo {author} {\bibfnamefont {D.~E.~P.}\ \bibnamefont {Vanpoucke}}, \bibinfo {author} {\bibfnamefont {S.}~\bibnamefont {Vijay}}, \bibinfo {author} {\bibfnamefont {M.}~\bibnamefont {Wolloch}}, \bibinfo {author} {\bibfnamefont {D.}~\bibnamefont {Wortmann}}, \bibinfo {author} {\bibfnamefont {A.~V.}\ \bibnamefont {Yakutovich}}, \bibinfo {author} {\bibfnamefont {J.}~\bibnamefont {Yu}}, \bibinfo {author} {\bibfnamefont {A.}~\bibnamefont {Zadoks}}, \bibinfo {author} {\bibfnamefont {B.}~\bibnamefont {Zhu}}, \ and\ \bibinfo {author} {\bibfnamefont {G.}~\bibnamefont {Pizzi}},\ }\href {\doibase 10.1038/s42254-023-00655-3} {\bibfield  {journal} {\bibinfo  {journal} {Nature Reviews Physics}\ }\textbf {\bibinfo {volume} {6}},\ \bibinfo {pages} {45} (\bibinfo
  {year} {2024})}\BibitemShut {NoStop}%
\bibitem [{\citenamefont {Onida}\ \emph {et~al.}(2002)\citenamefont {Onida}, \citenamefont {Reining},\ and\ \citenamefont {Rubio}}]{OnidaElectronic2002}%
  \BibitemOpen
  \bibfield  {author} {\bibinfo {author} {\bibfnamefont {G.}~\bibnamefont {Onida}}, \bibinfo {author} {\bibfnamefont {L.}~\bibnamefont {Reining}}, \ and\ \bibinfo {author} {\bibfnamefont {A.}~\bibnamefont {Rubio}},\ }\href {\doibase 10.1103/RevModPhys.74.601} {\bibfield  {journal} {\bibinfo  {journal} {Rev. Mod. Phys.}\ }\textbf {\bibinfo {volume} {74}},\ \bibinfo {pages} {601} (\bibinfo {year} {2002})}\BibitemShut {NoStop}%
\bibitem [{\citenamefont {van Schilfgaarde}\ \emph {et~al.}(2006)\citenamefont {van Schilfgaarde}, \citenamefont {Kotani},\ and\ \citenamefont {Faleev}}]{SchilfgaardeAdequacy2006}%
  \BibitemOpen
  \bibfield  {author} {\bibinfo {author} {\bibfnamefont {M.}~\bibnamefont {van Schilfgaarde}}, \bibinfo {author} {\bibfnamefont {T.}~\bibnamefont {Kotani}}, \ and\ \bibinfo {author} {\bibfnamefont {S.~V.}\ \bibnamefont {Faleev}},\ }\href {\doibase 10.1103/PhysRevB.74.245125} {\bibfield  {journal} {\bibinfo  {journal} {Phys. Rev. B}\ }\textbf {\bibinfo {volume} {74}},\ \bibinfo {pages} {245125} (\bibinfo {year} {2006})}\BibitemShut {NoStop}%
\bibitem [{\citenamefont {G\'omez-Abal}\ \emph {et~al.}(2008)\citenamefont {G\'omez-Abal}, \citenamefont {Li}, \citenamefont {Scheffler},\ and\ \citenamefont {Ambrosch-Draxl}}]{GomezInfluence2008}%
  \BibitemOpen
  \bibfield  {author} {\bibinfo {author} {\bibfnamefont {R.}~\bibnamefont {G\'omez-Abal}}, \bibinfo {author} {\bibfnamefont {X.}~\bibnamefont {Li}}, \bibinfo {author} {\bibfnamefont {M.}~\bibnamefont {Scheffler}}, \ and\ \bibinfo {author} {\bibfnamefont {C.}~\bibnamefont {Ambrosch-Draxl}},\ }\href {\doibase 10.1103/PhysRevLett.101.106404} {\bibfield  {journal} {\bibinfo  {journal} {Phys. Rev. Lett.}\ }\textbf {\bibinfo {volume} {101}},\ \bibinfo {pages} {106404} (\bibinfo {year} {2008})}\BibitemShut {NoStop}%
\bibitem [{\citenamefont {Klime\ifmmode~\check{s}\else \v{s}\fi{}}\ \emph {et~al.}(2014)\citenamefont {Klime\ifmmode~\check{s}\else \v{s}\fi{}}, \citenamefont {Kaltak},\ and\ \citenamefont {Kresse}}]{KlimesPredictive2014}%
  \BibitemOpen
  \bibfield  {author} {\bibinfo {author} {\bibfnamefont {J.~c.~v.}\ \bibnamefont {Klime\ifmmode~\check{s}\else \v{s}\fi{}}}, \bibinfo {author} {\bibfnamefont {M.}~\bibnamefont {Kaltak}}, \ and\ \bibinfo {author} {\bibfnamefont {G.}~\bibnamefont {Kresse}},\ }\href {\doibase 10.1103/PhysRevB.90.075125} {\bibfield  {journal} {\bibinfo  {journal} {Phys. Rev. B}\ }\textbf {\bibinfo {volume} {90}},\ \bibinfo {pages} {075125} (\bibinfo {year} {2014})}\BibitemShut {NoStop}%
\bibitem [{\citenamefont {Govoni}\ and\ \citenamefont {Galli}(2018)}]{GovoniGW1002018}%
  \BibitemOpen
  \bibfield  {author} {\bibinfo {author} {\bibfnamefont {M.}~\bibnamefont {Govoni}}\ and\ \bibinfo {author} {\bibfnamefont {G.}~\bibnamefont {Galli}},\ }\href {\doibase 10.1021/acs.jctc.7b00952} {\bibfield  {journal} {\bibinfo  {journal} {J. Chem. Theory Comput.}\ }\textbf {\bibinfo {volume} {14}},\ \bibinfo {pages} {1895} (\bibinfo {year} {2018})}\BibitemShut {NoStop}%
\bibitem [{\citenamefont {Garrity}\ \emph {et~al.}(2014)\citenamefont {Garrity}, \citenamefont {Bennett}, \citenamefont {Rabe},\ and\ \citenamefont {Vanderbilt}}]{GarrityPseudopotentials2014}%
  \BibitemOpen
  \bibfield  {author} {\bibinfo {author} {\bibfnamefont {K.~F.}\ \bibnamefont {Garrity}}, \bibinfo {author} {\bibfnamefont {J.~W.}\ \bibnamefont {Bennett}}, \bibinfo {author} {\bibfnamefont {K.~M.}\ \bibnamefont {Rabe}}, \ and\ \bibinfo {author} {\bibfnamefont {D.}~\bibnamefont {Vanderbilt}},\ }\href {\doibase https://doi.org/10.1016/j.commatsci.2013.08.053} {\bibfield  {journal} {\bibinfo  {journal} {Computational Materials Science}\ }\textbf {\bibinfo {volume} {81}},\ \bibinfo {pages} {446} (\bibinfo {year} {2014})}\BibitemShut {NoStop}%
\bibitem [{\citenamefont {{van Setten}}\ \emph {et~al.}(2018)\citenamefont {{van Setten}}, \citenamefont {Giantomassi}, \citenamefont {Bousquet}, \citenamefont {Verstraete}, \citenamefont {Hamann}, \citenamefont {Gonze},\ and\ \citenamefont {Rignanese}}]{vanSettenThePseudoDojo2018}%
  \BibitemOpen
  \bibfield  {author} {\bibinfo {author} {\bibfnamefont {M.}~\bibnamefont {{van Setten}}}, \bibinfo {author} {\bibfnamefont {M.}~\bibnamefont {Giantomassi}}, \bibinfo {author} {\bibfnamefont {E.}~\bibnamefont {Bousquet}}, \bibinfo {author} {\bibfnamefont {M.}~\bibnamefont {Verstraete}}, \bibinfo {author} {\bibfnamefont {D.}~\bibnamefont {Hamann}}, \bibinfo {author} {\bibfnamefont {X.}~\bibnamefont {Gonze}}, \ and\ \bibinfo {author} {\bibfnamefont {G.-M.}\ \bibnamefont {Rignanese}},\ }\href {\doibase https://doi.org/10.1016/j.cpc.2018.01.012} {\bibfield  {journal} {\bibinfo  {journal} {Computer Physics Communications}\ }\textbf {\bibinfo {volume} {226}},\ \bibinfo {pages} {39} (\bibinfo {year} {2018})}\BibitemShut {NoStop}%
\bibitem [{\citenamefont {Azizi}\ \emph {et~al.}(2025)\citenamefont {Azizi}, \citenamefont {Delesma}, \citenamefont {Giantomassi}, \citenamefont {Zavickis}, \citenamefont {Kuisma}, \citenamefont {Thyghesen}, \citenamefont {Golze}, \citenamefont {Buccheri}, \citenamefont {Zhang}, \citenamefont {Rinke}, \citenamefont {Draxl}, \citenamefont {Gulans},\ and\ \citenamefont {Gonze}}]{AziziPrecision2025}%
  \BibitemOpen
  \bibfield  {author} {\bibinfo {author} {\bibfnamefont {M.}~\bibnamefont {Azizi}}, \bibinfo {author} {\bibfnamefont {F.~A.}\ \bibnamefont {Delesma}}, \bibinfo {author} {\bibfnamefont {M.}~\bibnamefont {Giantomassi}}, \bibinfo {author} {\bibfnamefont {D.}~\bibnamefont {Zavickis}}, \bibinfo {author} {\bibfnamefont {M.}~\bibnamefont {Kuisma}}, \bibinfo {author} {\bibfnamefont {K.}~\bibnamefont {Thyghesen}}, \bibinfo {author} {\bibfnamefont {D.}~\bibnamefont {Golze}}, \bibinfo {author} {\bibfnamefont {A.}~\bibnamefont {Buccheri}}, \bibinfo {author} {\bibfnamefont {M.-Y.}\ \bibnamefont {Zhang}}, \bibinfo {author} {\bibfnamefont {P.}~\bibnamefont {Rinke}}, \bibinfo {author} {\bibfnamefont {C.}~\bibnamefont {Draxl}}, \bibinfo {author} {\bibfnamefont {A.}~\bibnamefont {Gulans}}, \ and\ \bibinfo {author} {\bibfnamefont {X.}~\bibnamefont {Gonze}},\ }\href {\doibase https://doi.org/10.1016/j.commatsci.2024.113655} {\bibfield  {journal} {\bibinfo  {journal} {Computational Materials Science}\ }\textbf {\bibinfo {volume}
  {250}},\ \bibinfo {pages} {113655} (\bibinfo {year} {2025})}\BibitemShut {NoStop}%
\bibitem [{\citenamefont {Bl\"ochl}(1994)}]{BlochlProjector1994}%
  \BibitemOpen
  \bibfield  {author} {\bibinfo {author} {\bibfnamefont {P.~E.}\ \bibnamefont {Bl\"ochl}},\ }\href {\doibase 10.1103/PhysRevB.50.17953} {\bibfield  {journal} {\bibinfo  {journal} {Phys. Rev. B}\ }\textbf {\bibinfo {volume} {50}},\ \bibinfo {pages} {17953} (\bibinfo {year} {1994})}\BibitemShut {NoStop}%
\bibitem [{\citenamefont {Kresse}\ and\ \citenamefont {Joubert}(1999)}]{KressefromUltrasoft1999}%
  \BibitemOpen
  \bibfield  {author} {\bibinfo {author} {\bibfnamefont {G.}~\bibnamefont {Kresse}}\ and\ \bibinfo {author} {\bibfnamefont {D.}~\bibnamefont {Joubert}},\ }\href {\doibase 10.1103/PhysRevB.59.1758} {\bibfield  {journal} {\bibinfo  {journal} {Phys. Rev. B}\ }\textbf {\bibinfo {volume} {59}},\ \bibinfo {pages} {1758} (\bibinfo {year} {1999})}\BibitemShut {NoStop}%
\bibitem [{\citenamefont {Gr\"uneis}\ \emph {et~al.}(2014)\citenamefont {Gr\"uneis}, \citenamefont {Kresse}, \citenamefont {Hinuma},\ and\ \citenamefont {Oba}}]{GruneisIonization2014}%
  \BibitemOpen
  \bibfield  {author} {\bibinfo {author} {\bibfnamefont {A.}~\bibnamefont {Gr\"uneis}}, \bibinfo {author} {\bibfnamefont {G.}~\bibnamefont {Kresse}}, \bibinfo {author} {\bibfnamefont {Y.}~\bibnamefont {Hinuma}}, \ and\ \bibinfo {author} {\bibfnamefont {F.}~\bibnamefont {Oba}},\ }\href {\doibase 10.1103/PhysRevLett.112.096401} {\bibfield  {journal} {\bibinfo  {journal} {Phys. Rev. Lett.}\ }\textbf {\bibinfo {volume} {112}},\ \bibinfo {pages} {096401} (\bibinfo {year} {2014})}\BibitemShut {NoStop}%
\bibitem [{\citenamefont {Georges}\ \emph {et~al.}(1996)\citenamefont {Georges}, \citenamefont {Kotliar}, \citenamefont {Krauth},\ and\ \citenamefont {Rozenberg}}]{Georges1996Dynamical}%
  \BibitemOpen
  \bibfield  {author} {\bibinfo {author} {\bibfnamefont {A.}~\bibnamefont {Georges}}, \bibinfo {author} {\bibfnamefont {G.}~\bibnamefont {Kotliar}}, \bibinfo {author} {\bibfnamefont {W.}~\bibnamefont {Krauth}}, \ and\ \bibinfo {author} {\bibfnamefont {M.~J.}\ \bibnamefont {Rozenberg}},\ }\href {\doibase 10.1103/RevModPhys.68.13} {\bibfield  {journal} {\bibinfo  {journal} {Rev. Mod. Phys.}\ }\textbf {\bibinfo {volume} {68}},\ \bibinfo {pages} {13} (\bibinfo {year} {1996})}\BibitemShut {NoStop}%
\bibitem [{\citenamefont {Chiarotti}\ \emph {et~al.}(2024)\citenamefont {Chiarotti}, \citenamefont {Ferretti},\ and\ \citenamefont {Marzari}}]{Chiarotti2024Energies}%
  \BibitemOpen
  \bibfield  {author} {\bibinfo {author} {\bibfnamefont {T.}~\bibnamefont {Chiarotti}}, \bibinfo {author} {\bibfnamefont {A.}~\bibnamefont {Ferretti}}, \ and\ \bibinfo {author} {\bibfnamefont {N.}~\bibnamefont {Marzari}},\ }\href {\doibase 10.1103/PhysRevResearch.6.L032023} {\bibfield  {journal} {\bibinfo  {journal} {Phys. Rev. Res.}\ }\textbf {\bibinfo {volume} {6}},\ \bibinfo {pages} {L032023} (\bibinfo {year} {2024})}\BibitemShut {NoStop}%
\bibitem [{\citenamefont {Chiarotti}\ \emph {et~al.}(2025)\citenamefont {Chiarotti}, \citenamefont {Quinzi}, \citenamefont {Pintus}, \citenamefont {Caserta}, \citenamefont {Ferretti},\ and\ \citenamefont {Marzari}}]{Chiarotti2025selfconsistent}%
  \BibitemOpen
  \bibfield  {author} {\bibinfo {author} {\bibfnamefont {T.}~\bibnamefont {Chiarotti}}, \bibinfo {author} {\bibfnamefont {M.}~\bibnamefont {Quinzi}}, \bibinfo {author} {\bibfnamefont {A.}~\bibnamefont {Pintus}}, \bibinfo {author} {\bibfnamefont {M.}~\bibnamefont {Caserta}}, \bibinfo {author} {\bibfnamefont {A.}~\bibnamefont {Ferretti}}, \ and\ \bibinfo {author} {\bibfnamefont {N.}~\bibnamefont {Marzari}},\ }\href {https://arxiv.org/abs/2508.18194} {} (\bibinfo {year} {2025}),\ \Eprint {http://arxiv.org/abs/2508.18194} {arXiv:2508.18194 [cond-mat.str-el]} \BibitemShut {NoStop}%
\bibitem [{\citenamefont {Chiarotti}\ \emph {et~al.}(2022)\citenamefont {Chiarotti}, \citenamefont {Marzari},\ and\ \citenamefont {Ferretti}}]{Chiarotti2022Unified}%
  \BibitemOpen
  \bibfield  {author} {\bibinfo {author} {\bibfnamefont {T.}~\bibnamefont {Chiarotti}}, \bibinfo {author} {\bibfnamefont {N.}~\bibnamefont {Marzari}}, \ and\ \bibinfo {author} {\bibfnamefont {A.}~\bibnamefont {Ferretti}},\ }\href {\doibase 10.1103/PhysRevResearch.4.013242} {\bibfield  {journal} {\bibinfo  {journal} {Phys. Rev. Res.}\ }\textbf {\bibinfo {volume} {4}},\ \bibinfo {pages} {013242} (\bibinfo {year} {2022})}\BibitemShut {NoStop}%
\bibitem [{Note1()}]{Note1}%
  \BibitemOpen
  \bibinfo {note} {A braket notation is used even for non-normalizable states, in agreement with Refs.~\cite {VanderbiltSoft1990, HamannOptimized2013}. All integrals are still meaningful as they involve wavefunctions up to a finite cutoff radius $R_c$.}\BibitemShut {Stop}%
\bibitem [{\citenamefont {Hamann}(1989)}]{Hamann1989Generalized}%
  \BibitemOpen
  \bibfield  {author} {\bibinfo {author} {\bibfnamefont {D.~R.}\ \bibnamefont {Hamann}},\ }\href {\doibase 10.1103/PhysRevB.40.2980} {\bibfield  {journal} {\bibinfo  {journal} {Phys. Rev. B}\ }\textbf {\bibinfo {volume} {40}},\ \bibinfo {pages} {2980} (\bibinfo {year} {1989})}\BibitemShut {NoStop}%
\bibitem [{\citenamefont {Martin}\ \emph {et~al.}(2016)\citenamefont {Martin}, \citenamefont {Reining},\ and\ \citenamefont {Ceperley}}]{Martin_Reining_Ceperley_2016}%
  \BibitemOpen
  \bibfield  {author} {\bibinfo {author} {\bibfnamefont {R.~M.}\ \bibnamefont {Martin}}, \bibinfo {author} {\bibfnamefont {L.}~\bibnamefont {Reining}}, \ and\ \bibinfo {author} {\bibfnamefont {D.~M.}\ \bibnamefont {Ceperley}},\ }\href@noop {} {\emph {\bibinfo {title} {Interacting Electrons: Theory and Computational Approaches}}}\ (\bibinfo  {publisher} {Cambridge University Press},\ \bibinfo {year} {2016})\BibitemShut {NoStop}%
\bibitem [{\citenamefont {Martin}(2020)}]{Martin_2020}%
  \BibitemOpen
  \bibfield  {author} {\bibinfo {author} {\bibfnamefont {R.~M.}\ \bibnamefont {Martin}},\ }\href@noop {} {\emph {\bibinfo {title} {Electronic Structure: Basic Theory and Practical Methods}}},\ \bibinfo {edition} {2nd}\ ed.\ (\bibinfo  {publisher} {Cambridge University Press},\ \bibinfo {year} {2020})\BibitemShut {NoStop}%
\bibitem [{\citenamefont {L\"uders}(1955)}]{Luders1955Zum}%
  \BibitemOpen
  \bibfield  {author} {\bibinfo {author} {\bibfnamefont {G.}~\bibnamefont {L\"uders}},\ }\href@noop {} {\bibfield  {journal} {\bibinfo  {journal} {Z. Naturforsch.}\ }\textbf {\bibinfo {volume} {10a}},\ \bibinfo {pages} {581} (\bibinfo {year} {1955})}\BibitemShut {NoStop}%
\bibitem [{\citenamefont {Shirley}\ \emph {et~al.}(1989)\citenamefont {Shirley}, \citenamefont {Allan}, \citenamefont {Martin},\ and\ \citenamefont {Joannopoulos}}]{Shirley1989Extended}%
  \BibitemOpen
  \bibfield  {author} {\bibinfo {author} {\bibfnamefont {E.~L.}\ \bibnamefont {Shirley}}, \bibinfo {author} {\bibfnamefont {D.~C.}\ \bibnamefont {Allan}}, \bibinfo {author} {\bibfnamefont {R.~M.}\ \bibnamefont {Martin}}, \ and\ \bibinfo {author} {\bibfnamefont {J.~D.}\ \bibnamefont {Joannopoulos}},\ }\href {\doibase 10.1103/PhysRevB.40.3652} {\bibfield  {journal} {\bibinfo  {journal} {Phys. Rev. B}\ }\textbf {\bibinfo {volume} {40}},\ \bibinfo {pages} {3652} (\bibinfo {year} {1989})}\BibitemShut {NoStop}%
\bibitem [{\citenamefont {Shirley}\ and\ \citenamefont {Martin}(1993)}]{Shirley1993Many-body}%
  \BibitemOpen
  \bibfield  {author} {\bibinfo {author} {\bibfnamefont {E.~L.}\ \bibnamefont {Shirley}}\ and\ \bibinfo {author} {\bibfnamefont {R.~M.}\ \bibnamefont {Martin}},\ }\href {\doibase 10.1103/PhysRevB.47.15413} {\bibfield  {journal} {\bibinfo  {journal} {Phys. Rev. B}\ }\textbf {\bibinfo {volume} {47}},\ \bibinfo {pages} {15413} (\bibinfo {year} {1993})}\BibitemShut {NoStop}%
\bibitem [{\citenamefont {Stefanucci}\ and\ \citenamefont {van Leeuwen}(2013)}]{Stefanucci_vanLeeuwen_2013}%
  \BibitemOpen
  \bibfield  {author} {\bibinfo {author} {\bibfnamefont {G.}~\bibnamefont {Stefanucci}}\ and\ \bibinfo {author} {\bibfnamefont {R.}~\bibnamefont {van Leeuwen}},\ }\href@noop {} {\emph {\bibinfo {title} {Nonequilibrium Many-Body Theory of Quantum Systems: A Modern Introduction}}}\ (\bibinfo  {publisher} {Cambridge University Press},\ \bibinfo {year} {2013})\BibitemShut {NoStop}%
\bibitem [{\citenamefont {Ferretti}\ \emph {et~al.}(2024)\citenamefont {Ferretti}, \citenamefont {Chiarotti},\ and\ \citenamefont {Marzari}}]{Ferretti2024Greens}%
  \BibitemOpen
  \bibfield  {author} {\bibinfo {author} {\bibfnamefont {A.}~\bibnamefont {Ferretti}}, \bibinfo {author} {\bibfnamefont {T.}~\bibnamefont {Chiarotti}}, \ and\ \bibinfo {author} {\bibfnamefont {N.}~\bibnamefont {Marzari}},\ }\href {\doibase 10.1103/PhysRevB.110.045149} {\bibfield  {journal} {\bibinfo  {journal} {Phys. Rev. B}\ }\textbf {\bibinfo {volume} {110}},\ \bibinfo {pages} {045149} (\bibinfo {year} {2024})}\BibitemShut {NoStop}%
\bibitem [{\citenamefont {Marzari}\ and\ \citenamefont {Vanderbilt}(1997)}]{Marzari1997Maximally}%
  \BibitemOpen
  \bibfield  {author} {\bibinfo {author} {\bibfnamefont {N.}~\bibnamefont {Marzari}}\ and\ \bibinfo {author} {\bibfnamefont {D.}~\bibnamefont {Vanderbilt}},\ }\href {\doibase 10.1103/PhysRevB.56.12847} {\bibfield  {journal} {\bibinfo  {journal} {Phys. Rev. B}\ }\textbf {\bibinfo {volume} {56}},\ \bibinfo {pages} {12847} (\bibinfo {year} {1997})}\BibitemShut {NoStop}%
\bibitem [{\citenamefont {Perdew}\ \emph {et~al.}(1996)\citenamefont {Perdew}, \citenamefont {Burke},\ and\ \citenamefont {Ernzerhof}}]{PBE_PhysRevLett.77.3865_1996}%
  \BibitemOpen
  \bibfield  {author} {\bibinfo {author} {\bibfnamefont {J.~P.}\ \bibnamefont {Perdew}}, \bibinfo {author} {\bibfnamefont {K.}~\bibnamefont {Burke}}, \ and\ \bibinfo {author} {\bibfnamefont {M.}~\bibnamefont {Ernzerhof}},\ }\href {\doibase 10.1103/PhysRevLett.77.3865} {\bibfield  {journal} {\bibinfo  {journal} {Phys. Rev. Lett.}\ }\textbf {\bibinfo {volume} {77}},\ \bibinfo {pages} {3865} (\bibinfo {year} {1996})}\BibitemShut {NoStop}%
\bibitem [{\citenamefont {Gonze}\ \emph {et~al.}(1991)\citenamefont {Gonze}, \citenamefont {Stumpf},\ and\ \citenamefont {Scheffler}}]{Gonze1991Analysis}%
  \BibitemOpen
  \bibfield  {author} {\bibinfo {author} {\bibfnamefont {X.}~\bibnamefont {Gonze}}, \bibinfo {author} {\bibfnamefont {R.}~\bibnamefont {Stumpf}}, \ and\ \bibinfo {author} {\bibfnamefont {M.}~\bibnamefont {Scheffler}},\ }\href {\doibase 10.1103/PhysRevB.44.8503} {\bibfield  {journal} {\bibinfo  {journal} {Phys. Rev. B}\ }\textbf {\bibinfo {volume} {44}},\ \bibinfo {pages} {8503} (\bibinfo {year} {1991})}\BibitemShut {NoStop}%
\bibitem [{\citenamefont {Prandini}\ \emph {et~al.}(2018)\citenamefont {Prandini}, \citenamefont {Marrazzo}, \citenamefont {Castelli}, \citenamefont {Mounet},\ and\ \citenamefont {Marzari}}]{Prandini2018Precision}%
  \BibitemOpen
  \bibfield  {author} {\bibinfo {author} {\bibfnamefont {G.}~\bibnamefont {Prandini}}, \bibinfo {author} {\bibfnamefont {A.}~\bibnamefont {Marrazzo}}, \bibinfo {author} {\bibfnamefont {I.~E.}\ \bibnamefont {Castelli}}, \bibinfo {author} {\bibfnamefont {N.}~\bibnamefont {Mounet}}, \ and\ \bibinfo {author} {\bibfnamefont {N.}~\bibnamefont {Marzari}},\ }\href {\doibase 10.1038/s41524-018-0127-2} {\bibfield  {journal} {\bibinfo  {journal} {npj Computational Materials}\ }\textbf {\bibinfo {volume} {4}},\ \bibinfo {pages} {72} (\bibinfo {year} {2018})}\BibitemShut {NoStop}%
\bibitem [{\citenamefont {Louie}\ \emph {et~al.}(1982)\citenamefont {Louie}, \citenamefont {Froyen},\ and\ \citenamefont {Cohen}}]{Louie1982Nonlinear}%
  \BibitemOpen
  \bibfield  {author} {\bibinfo {author} {\bibfnamefont {S.~G.}\ \bibnamefont {Louie}}, \bibinfo {author} {\bibfnamefont {S.}~\bibnamefont {Froyen}}, \ and\ \bibinfo {author} {\bibfnamefont {M.~L.}\ \bibnamefont {Cohen}},\ }\href {\doibase 10.1103/PhysRevB.26.1738} {\bibfield  {journal} {\bibinfo  {journal} {Phys. Rev. B}\ }\textbf {\bibinfo {volume} {26}},\ \bibinfo {pages} {1738} (\bibinfo {year} {1982})}\BibitemShut {NoStop}%
\bibitem [{\citenamefont {Luttinger}\ and\ \citenamefont {Ward}(1960)}]{Luttinger1960Ground-state}%
  \BibitemOpen
  \bibfield  {author} {\bibinfo {author} {\bibfnamefont {J.~M.}\ \bibnamefont {Luttinger}}\ and\ \bibinfo {author} {\bibfnamefont {J.~C.}\ \bibnamefont {Ward}},\ }\href {\doibase 10.1103/PhysRev.118.1417} {\bibfield  {journal} {\bibinfo  {journal} {Phys. Rev.}\ }\textbf {\bibinfo {volume} {118}},\ \bibinfo {pages} {1417} (\bibinfo {year} {1960})}\BibitemShut {NoStop}%
\bibitem [{\citenamefont {Klein}(1961)}]{Klein1961Perturbation}%
  \BibitemOpen
  \bibfield  {author} {\bibinfo {author} {\bibfnamefont {A.}~\bibnamefont {Klein}},\ }\href {\doibase 10.1103/PhysRev.121.950} {\bibfield  {journal} {\bibinfo  {journal} {Phys. Rev.}\ }\textbf {\bibinfo {volume} {121}},\ \bibinfo {pages} {950} (\bibinfo {year} {1961})}\BibitemShut {NoStop}%
\bibitem [{\citenamefont {Baym}\ and\ \citenamefont {Kadanoff}(1961)}]{Baym1961Conservation}%
  \BibitemOpen
  \bibfield  {author} {\bibinfo {author} {\bibfnamefont {G.}~\bibnamefont {Baym}}\ and\ \bibinfo {author} {\bibfnamefont {L.~P.}\ \bibnamefont {Kadanoff}},\ }\href {\doibase 10.1103/PhysRev.124.287} {\bibfield  {journal} {\bibinfo  {journal} {Phys. Rev.}\ }\textbf {\bibinfo {volume} {124}},\ \bibinfo {pages} {287} (\bibinfo {year} {1961})}\BibitemShut {NoStop}%
\bibitem [{\citenamefont {Almbladh}\ \emph {et~al.}(1999)\citenamefont {Almbladh}, \citenamefont {Barth},\ and\ \citenamefont {Leeuwen}}]{Almbladh1999Variational}%
  \BibitemOpen
  \bibfield  {author} {\bibinfo {author} {\bibfnamefont {C.~O.}\ \bibnamefont {Almbladh}}, \bibinfo {author} {\bibfnamefont {U.~V.}\ \bibnamefont {Barth}}, \ and\ \bibinfo {author} {\bibfnamefont {R.~V.}\ \bibnamefont {Leeuwen}},\ }\href {\doibase 10.1142/S0217979299000436} {\bibfield  {journal} {\bibinfo  {journal} {International Journal of Modern Physics B}\ }\textbf {\bibinfo {volume} {13}},\ \bibinfo {pages} {535} (\bibinfo {year} {1999})}\BibitemShut {NoStop}%
\bibitem [{\citenamefont {Ferretti}\ and\ \citenamefont {Marzari}(2025)}]{Ferretti2025Functional}%
  \BibitemOpen
  \bibfield  {author} {\bibinfo {author} {\bibfnamefont {A.}~\bibnamefont {Ferretti}}\ and\ \bibinfo {author} {\bibfnamefont {N.}~\bibnamefont {Marzari}},\ }\href {https://arxiv.org/abs/2508.17245} {} (\bibinfo {year} {2025}),\ \Eprint {http://arxiv.org/abs/2508.17245} {arXiv:2508.17245 [cond-mat.mtrl-sci]} \BibitemShut {NoStop}%
\bibitem [{\citenamefont {Laasonen}\ \emph {et~al.}(1991)\citenamefont {Laasonen}, \citenamefont {Car}, \citenamefont {Lee},\ and\ \citenamefont {Vanderbilt}}]{Laasonen1991Implementation}%
  \BibitemOpen
  \bibfield  {author} {\bibinfo {author} {\bibfnamefont {K.}~\bibnamefont {Laasonen}}, \bibinfo {author} {\bibfnamefont {R.}~\bibnamefont {Car}}, \bibinfo {author} {\bibfnamefont {C.}~\bibnamefont {Lee}}, \ and\ \bibinfo {author} {\bibfnamefont {D.}~\bibnamefont {Vanderbilt}},\ }\href {\doibase 10.1103/PhysRevB.43.6796} {\bibfield  {journal} {\bibinfo  {journal} {Phys. Rev. B}\ }\textbf {\bibinfo {volume} {43}},\ \bibinfo {pages} {6796} (\bibinfo {year} {1991})}\BibitemShut {NoStop}%
\bibitem [{\citenamefont {Laasonen}\ \emph {et~al.}(1993)\citenamefont {Laasonen}, \citenamefont {Pasquarello}, \citenamefont {Car}, \citenamefont {Lee},\ and\ \citenamefont {Vanderbilt}}]{Laasonen1993Car-parrinello}%
  \BibitemOpen
  \bibfield  {author} {\bibinfo {author} {\bibfnamefont {K.}~\bibnamefont {Laasonen}}, \bibinfo {author} {\bibfnamefont {A.}~\bibnamefont {Pasquarello}}, \bibinfo {author} {\bibfnamefont {R.}~\bibnamefont {Car}}, \bibinfo {author} {\bibfnamefont {C.}~\bibnamefont {Lee}}, \ and\ \bibinfo {author} {\bibfnamefont {D.}~\bibnamefont {Vanderbilt}},\ }\href {\doibase 10.1103/PhysRevB.47.10142} {\bibfield  {journal} {\bibinfo  {journal} {Phys. Rev. B}\ }\textbf {\bibinfo {volume} {47}},\ \bibinfo {pages} {10142} (\bibinfo {year} {1993})}\BibitemShut {NoStop}%
\end{thebibliography}

\begin{thebibliography}{5}%
\makeatletter
\providecommand \@ifxundefined [1]{%
 \@ifx{#1\undefined}
}%
\providecommand \@ifnum [1]{%
 \ifnum #1\expandafter \@firstoftwo
 \else \expandafter \@secondoftwo
 \fi
}%
\providecommand \@ifx [1]{%
 \ifx #1\expandafter \@firstoftwo
 \else \expandafter \@secondoftwo
 \fi
}%
\providecommand \natexlab [1]{#1}%
\providecommand \enquote  [1]{``#1''}%
\providecommand \bibnamefont  [1]{#1}%
\providecommand \bibfnamefont [1]{#1}%
\providecommand \citenamefont [1]{#1}%
\providecommand \href@noop [0]{\@secondoftwo}%
\providecommand \href [0]{\begingroup \@sanitize@url \@href}%
\providecommand \@href[1]{\@@startlink{#1}\@@href}%
\providecommand \@@href[1]{\endgroup#1\@@endlink}%
\providecommand \@sanitize@url [0]{\catcode `\\12\catcode `\$12\catcode `\&12\catcode `\#12\catcode `\^12\catcode `\_12\catcode `\%12\relax}%
\providecommand \@@startlink[1]{}%
\providecommand \@@endlink[0]{}%
\providecommand \url  [0]{\begingroup\@sanitize@url \@url }%
\providecommand \@url [1]{\endgroup\@href {#1}{\urlprefix }}%
\providecommand \urlprefix  [0]{URL }%
\providecommand \Eprint [0]{\href }%
\providecommand \doibase [0]{https://doi.org/}%
\providecommand \selectlanguage [0]{\@gobble}%
\providecommand \bibinfo  [0]{\@secondoftwo}%
\providecommand \bibfield  [0]{\@secondoftwo}%
\providecommand \translation [1]{[#1]}%
\providecommand \BibitemOpen [0]{}%
\providecommand \bibitemStop [0]{}%
\providecommand \bibitemNoStop [0]{.\EOS\space}%
\providecommand \EOS [0]{\spacefactor3000\relax}%
\providecommand \BibitemShut  [1]{\csname bibitem#1\endcsname}%
\let\auto@bib@innerbib\@empty
\bibitem [{\citenamefont {Hamann}(1989)}]{Hamann1989GeneralizedSupp}%
  \BibitemOpen
  \bibfield  {author} {\bibinfo {author} {\bibfnamefont {D.~R.}\ \bibnamefont {Hamann}},\ }\href {https://doi.org/10.1103/PhysRevB.40.2980} {\bibfield  {journal} {\bibinfo  {journal} {Phys. Rev. B}\ }\textbf {\bibinfo {volume} {40}},\ \bibinfo {pages} {2980} (\bibinfo {year} {1989})}\BibitemShut {NoStop}%
\bibitem [{\citenamefont {Shirley}\ \emph {et~al.}(1989)\citenamefont {Shirley}, \citenamefont {Allan}, \citenamefont {Martin},\ and\ \citenamefont {Joannopoulos}}]{Shirley1989ExtendedSupp}%
  \BibitemOpen
  \bibfield  {author} {\bibinfo {author} {\bibfnamefont {E.~L.}\ \bibnamefont {Shirley}}, \bibinfo {author} {\bibfnamefont {D.~C.}\ \bibnamefont {Allan}}, \bibinfo {author} {\bibfnamefont {R.~M.}\ \bibnamefont {Martin}},\ and\ \bibinfo {author} {\bibfnamefont {J.~D.}\ \bibnamefont {Joannopoulos}},\ }\href {https://doi.org/10.1103/PhysRevB.40.3652} {\bibfield  {journal} {\bibinfo  {journal} {Phys. Rev. B}\ }\textbf {\bibinfo {volume} {40}},\ \bibinfo {pages} {3652} (\bibinfo {year} {1989})}\BibitemShut {NoStop}%
\bibitem [{\citenamefont {Martin}(2020)}]{Martin_2020Supp}%
  \BibitemOpen
  \bibfield  {author} {\bibinfo {author} {\bibfnamefont {R.~M.}\ \bibnamefont {Martin}},\ }\href@noop {} {\emph {\bibinfo {title} {Electronic Structure: Basic Theory and Practical Methods}}},\ \bibinfo {edition} {2nd}\ ed.\ (\bibinfo  {publisher} {Cambridge University Press},\ \bibinfo {year} {2020})\BibitemShut {NoStop}%
\bibitem [{\citenamefont {L\"uders}(1955)}]{Luders1955ZumSupp}%
  \BibitemOpen
  \bibfield  {author} {\bibinfo {author} {\bibfnamefont {G.}~\bibnamefont {L\"uders}},\ }\href@noop {} {\bibfield  {journal} {\bibinfo  {journal} {Z. Naturforsch.}\ }\textbf {\bibinfo {volume} {10a}},\ \bibinfo {pages} {581} (\bibinfo {year} {1955})}\BibitemShut {NoStop}%
\bibitem [{\citenamefont {Vanderbilt}(1990)}]{VanderbiltSoft1990Supp}%
  \BibitemOpen
  \bibfield  {author} {\bibinfo {author} {\bibfnamefont {D.}~\bibnamefont {Vanderbilt}},\ }\href {https://doi.org/10.1103/PhysRevB.41.7892} {\bibfield  {journal} {\bibinfo  {journal} {Phys. Rev. B}\ }\textbf {\bibinfo {volume} {41}},\ \bibinfo {pages} {7892} (\bibinfo {year} {1990})}\BibitemShut {NoStop}%
\end{thebibliography}
%
%
%
%
\end{document}